\documentclass[referee]{raa}           

\usepackage{graphicx,times}
\usepackage{natbib}
\usepackage{amssymb,amsmath}
\usepackage{listings}
\usepackage{verbatim}
\usepackage{microtype,siunitx,booktabs}
\usepackage{makecell}
\usepackage{array}
\usepackage{newtxtext,newtxmath}
\usepackage[T1]{fontenc}

\bibpunct{(}{)}{;}{a}{}{,}

\usepackage[pagebackref=true]{hyperref}

\begin{document}

\title{Comprehensive Gaia DR3-Based Astrometric, Photometric, and Kinematic Studies of the Binary Open Cluster $h$ and $\chi$ Persei}

 \volnopage{ {\bf 20XX} Vol.\ {\bf X} No. {\bf XX}, 000--000}
  \setcounter{page}{1}

\author{Seval Ta\c{s}demir\inst{1}
\and Waleed Elsanhoury\inst{2}
\and Deniz Cennet Çınar\inst{1}
\and Aly Haroon\inst{3,4},
\and Selçuk Bilir\inst{5,*}\footnotetext{$*$Corresponding author, \it sbilir@istanbul.edu.tr}
}

\institute{
Istanbul University, Institute of Graduate Studies in Science, Programme of Astronomy and Space Sciences, 34116, Beyazıt, Istanbul, Türkiye\\
\and
Northern Border University, College of Science, Department of Physics, Arar, Saudi Arabia\\
\and
King Abdulaziz University, Faculty of Science, Astronomy and Space Science Department, Jeddah, Kingdom of Saudi Arabia\\
\and
National Research Institute of Astronomy and Geophysics (NRIAG), Department of Astronomy, Cairo, Egypt\\
\and
Istanbul University, Faculty of Science, Department of Astronomy and Space Sciences, 34119, Beyazıt, Istanbul, Türkiye; {\it sbilir@istanbul.edu.tr}\\
}
\vs \no
   {\small accepted 2026 01 22}

\abstract{In the {\it Gaia} era, a comprehensive analysis of the binary open clusters NGC\, 869 ($h$ Persei) and NGC\,884 ($\chi$ Persei) system has been conducted to investigate its structural, astrophysical, kinematic, and Galactic orbital properties, along with its dynamical evolution. By applying the {\sc UPMASK} algorithm to {\it Gaia} astrometric data for the estimation of cluster membership probabilities, it has been determined that 808  stars in NGC 869 and 707  stars in NGC 884 exhibit the highest statistical likelihood of being cluster members. The fundamental astrophysical parameters of the clusters were inferred within a Bayesian framework using {\it Gaia} data and {\sc PARSEC} stellar evolutionary isochrones, through the application of the Markov Chain Monte Carlo (\texttt{MCMC}) technique. The estimated parameters are: colour excess $E(B-V)=0.516^{+0.17}_{-0.24}$ mag and $0.516^{+0.22}_{-0.33}$ mag, distances $2376^{+301}_{-278}$ and $2273^{+230}_{-290}$ pc, ages $\log(t/\mathrm{yr}) = 7.31^{+0.17}_{-0.32}$ and $\log(t/\mathrm{yr}) = 7.30^{+0.13}_{-0.29}$, and metallicities $[\mathrm{Fe}/\mathrm{H}]=-0.24\pm 0.12$ and \textbf{$[\mathrm{Fe}/\mathrm{H}]=-0.25\pm 0.12$}~dex for NGC\,869 and NGC\,884, respectively Since spectroscopic observations are not available for the clusters, SED analysis was employed for the member stars, yielding results consistent with those obtained using the \texttt{MCMC} method. Kinematic and Galactic orbital analyses suggest that the open clusters originated in nearby regions of the Galaxy. This interpretation is supported by their similar space velocities and Galactic orbital parameters. Furthermore, orbital integration over 1 Gyr indicates a potential interaction between the clusters within the next 11 Myr. This study provides strong evidence of a common origin and a possible future dynamical interaction, contributing valuable insights into the formation and evolution of binary open clusters in the Milky Way.}

\authorrunning{Ta\c{s}demir et al.} 

\titlerunning{Binary Open Cluster $h$ and $\chi$ Persei}  
  
\maketitle

\keywords{Galaxy: open cluster and associations: individual: NGC\,869 and NGC\,884, stars: Hertzsprung-Russell (HR) diagram, Galaxy: Stellar kinematics}

\section{Introduction}
Open clusters (OCs) are key tracers of the Galactic disc's structure, chemical evolution, and formation properties. They serve as fundamental laboratories for testing stellar evolution theories due to their relatively uniform ages and chemical compositions. While most OCs are found as single stellar groupings, a fraction appear in binary or multiple cluster systems, which can provide deeper insights into cluster formation mechanisms and dynamical evolution. Binary OCs (BOCs) offer key evidence that star clusters can form and evolve through interactions beyond isolated conditions. Based on a statistical analysis, \citet{Delafuente_2009} emphasized that the occurrence and properties of such pairs help constrain the models of clustered star formation and early dynamical evolution.

Recent studies have identified several OCs located nearby and potentially exhibiting mutual gravitational interactions \citep{Vereshchagin2022, Song2022, Li2024, Haroon2024, Palma2024, TasdemirCinar_2025}. These systems, referred to as close binary open clusters (CBOCs), are of particular interest as they provide unique astrophysical laboratories for understanding the formation, evolution, and disruption of stellar clusters. CBOCs not only allow researchers to trace the early conditions of star formation but also offer valuable insight into how clusters dynamically interact with each other and respond to the large-scale potential of the Milky Way (MW). As such, investigating these rare systems has become increasingly significant in the broader context of Galactic structure and star cluster evolution. BOCs are typically identified based on three primary criteria, namely spatial proximity, similar ages, coherent proper motions, and radial velocities. These systems are thought to originate from the fragmentation of the same giant molecular cloud \citep{Fujimoto1997} or through tidal capture processes \citep{Camargo2021}. BOCs studies allow us to explore whether such pairs are physically bound or merely line-of-sight coincidences.

NGC\,869 ($h$~Persei) and NGC\,884 ($\chi$~Persei), are two prominent and spatially close clusters that constitute the well-known BOC in Perseus. NGC\,869 and NGC\,884 are classified as Trumpler type I3r, indicating that they are rich clusters with a strong central concentration and a wide range of stellar magnitudes \citep{Ruprecht_1966}. The equatorial coordinates of NGC\,869 and NGC\,884 are $\alpha = 02^\mathrm{h}18^\mathrm{m}57^\mathrm{s}.84$, $\delta = +57^\circ08'02''.4$ and $\alpha = 02^\mathrm{h}22^\mathrm{m}20^\mathrm{s}.16$, $\delta = +57^\circ08'56''.4$ (J2000.0), respectively. These correspond to Galactic coordinates of $(l, b) = (134^\circ.6257$, -$3^\circ.7372)$ for NGC\,869 and ($135^\circ.0552$, -$3^\circ.5683)$ for NGC\,884. Also, NGC\,869 and NGC\,884 are well-known BOCs frequently referenced in the literature due to their spatial proximity and similar astrophysical properties. The advent of the \textit{Gaia} mission \citep{Gaia_mission}, with its unprecedented astrometric and photometric precision, has made it possible to re-examine these OCs in greater detail. Such high-quality data is essential for refining our understanding of their dynamical relationship, origin, and evolutionary status within the Galactic disc. The most prominent and widely studied CBOCs in the MW are the NGC\,869 and NGC\,884. They are located at a heliocentric distance of ~2.3-2.4 kpc, and both OCs exhibit remarkable similarities in age, kinematics, and stellar population characteristics \citep{Bragaglia2001, Currie2010, Cantat_2020}. Both ages are typically estimated between 12 and 15 Myr \citep{marco01}, and the OCs are among the most massive young OCs known in the MW.

In this study, we perform a photometric and kinematic analysis of NGC\,869 and NGC\,884 using {\it Gaia} DR3 data to investigate their structural parameters, membership, kinematics, and Galactic orbits, and to assess their physical association and formation history. Section \ref{Data} presents the astrometric and photometric data, while Section \ref{data_analyzis} outlines the methods and main results. Section \ref{Dynamic_Kinematics} discusses the clusters’ kinematical and dynamical properties, and Section \ref{Evolving_Times} examines their dynamical parameters and mass distributions. Finally, Section \ref{Discussion} discusses the obtained results in comparison with previous studies in the literature and summarises the main conclusions regarding the evolutionary status of both OCs.

\section{Data}\label{Data}

The photometric, astrometric, and spectroscopic data used in this study were obtained from the \textit{Gaia} DR3 catalogue \citep{Gaia_DR3}. The photometric data include $G$, $G_{\rm BP}$, and $G_{\rm RP}$ magnitudes, while the astrometric parameters comprise the equatorial coordinates ($\alpha, \delta$), trigonometric parallax ($\varpi$), and proper-motion components ($\mu_{\alpha} \cos \delta, \mu_{\delta}$). Additionally, radial velocity ($V_{\rm R}$) measurements derived from {\it Gaia} spectroscopic data were also used in the analysis, when available. All parameters, along with their associated uncertainties, were used in the analysis. 95,454 stellar sources were obtained from the \textit{Gaia} DR3 database within a 40 arcminute radius centred at $\alpha = 02^\mathrm{h}20^\mathrm{m}39^\mathrm{s}.12$, $\delta = +57^\circ08'31''.2$, encompassing the BOCs NGC 869 and NGC 884. Various filtering criteria were applied to increase the reliability of the data and to eliminate astrometrically problematic sources. \textcolor{black}{A total of 9,374 sources were eliminated from the dataset because they lacked a Renormalised Unit Weight Error (\texttt{RUWE}) value. According to \citet{Lindegren_2021}, sources with RUWE values around 1 demonstrate good agreement with the single-star model. In contrast, \texttt{RUWE} values of 1.4 \citep{Castro-Ginard2024} and above generally indicate binary stars or problematic sources for the astrometric solution. In this particular context, 2,975 stars with \texttt{RUWE} $\ge$ 1.4 were also eliminated due to their low accuracy. Among these, 1,829 stars were found to have $G$-band magnitudes brighter than $G=17$, suggesting that the astrometric quality may degrade significantly at fainter magnitudes. In addition, 517 stars bearing the \texttt{duplicate\_source} flag were eliminated from the dataset; 76 of these stars were subsequently identified as exhibiting high \texttt{RUWE} values. As a result of the applied filtering steps, a final, reliable sample of 92,038 stars was obtained and subsequently analysed in detail.}

\begin{figure}
\centering
\includegraphics[width=0.9\linewidth]{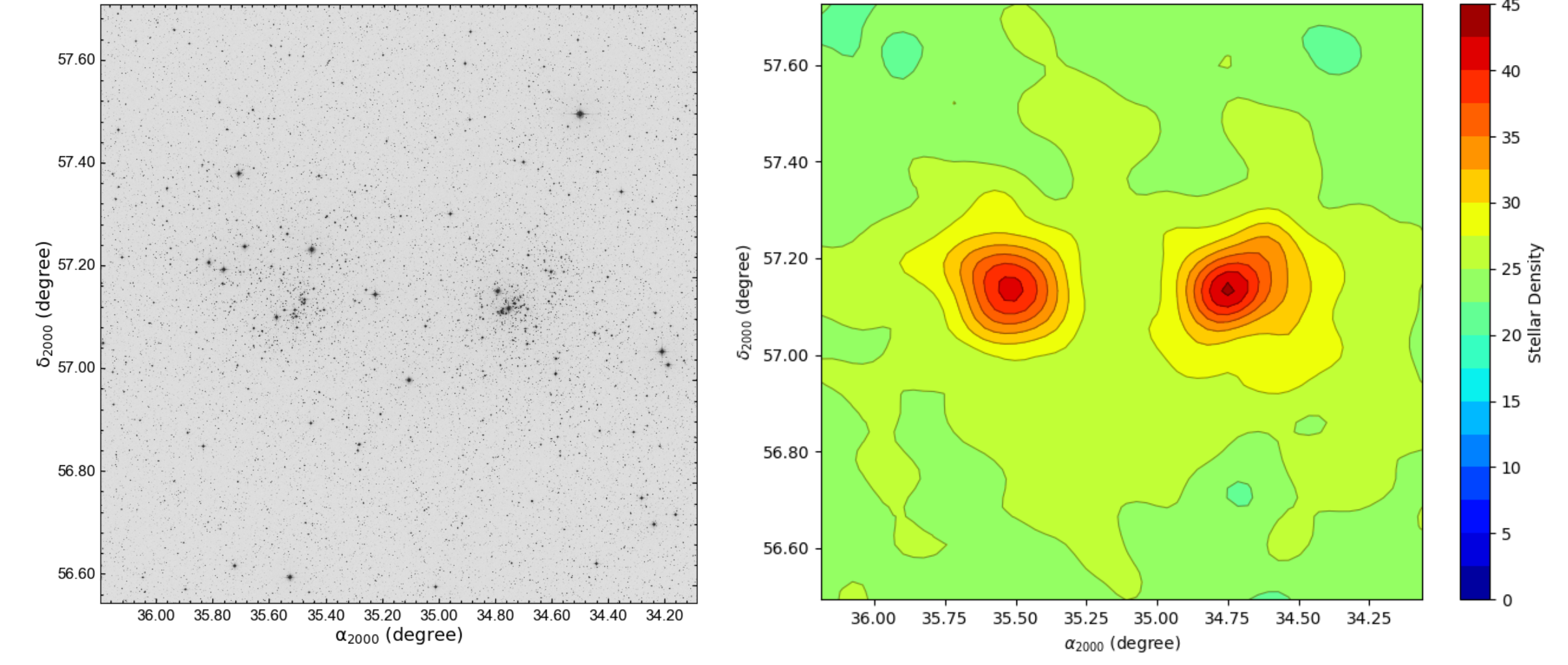}
\caption{Finder chart of the region containing the BOCs NGC 869 and NGC 884, obtained from the STScI Digitized Sky Survey (DSS). The red dashed circle indicates the area analyzed (left panel). Stellar surface density map of the same region, revealing two significant overdensities corresponding to NGC 869 (a) and NGC 884 (b) (right panel). Warmer colours represent higher stellar densities.}
\label{fig:chart}
\end{figure}

Additionally, an identification chart was created to provide a comprehensive spatial view of the system. This diagram utilizes the $80'\times 80'$ arcminutes of space in which both OCs are located, clearly showing the individual clusters and their positions relative to each other. The identification chart based on this field is demonstrated in Figure~\ref{fig:chart}, where NGC\,869 and NGC\,884 are distinctly visible.

\begin{figure}
\centering
\includegraphics[width=0.6\linewidth]{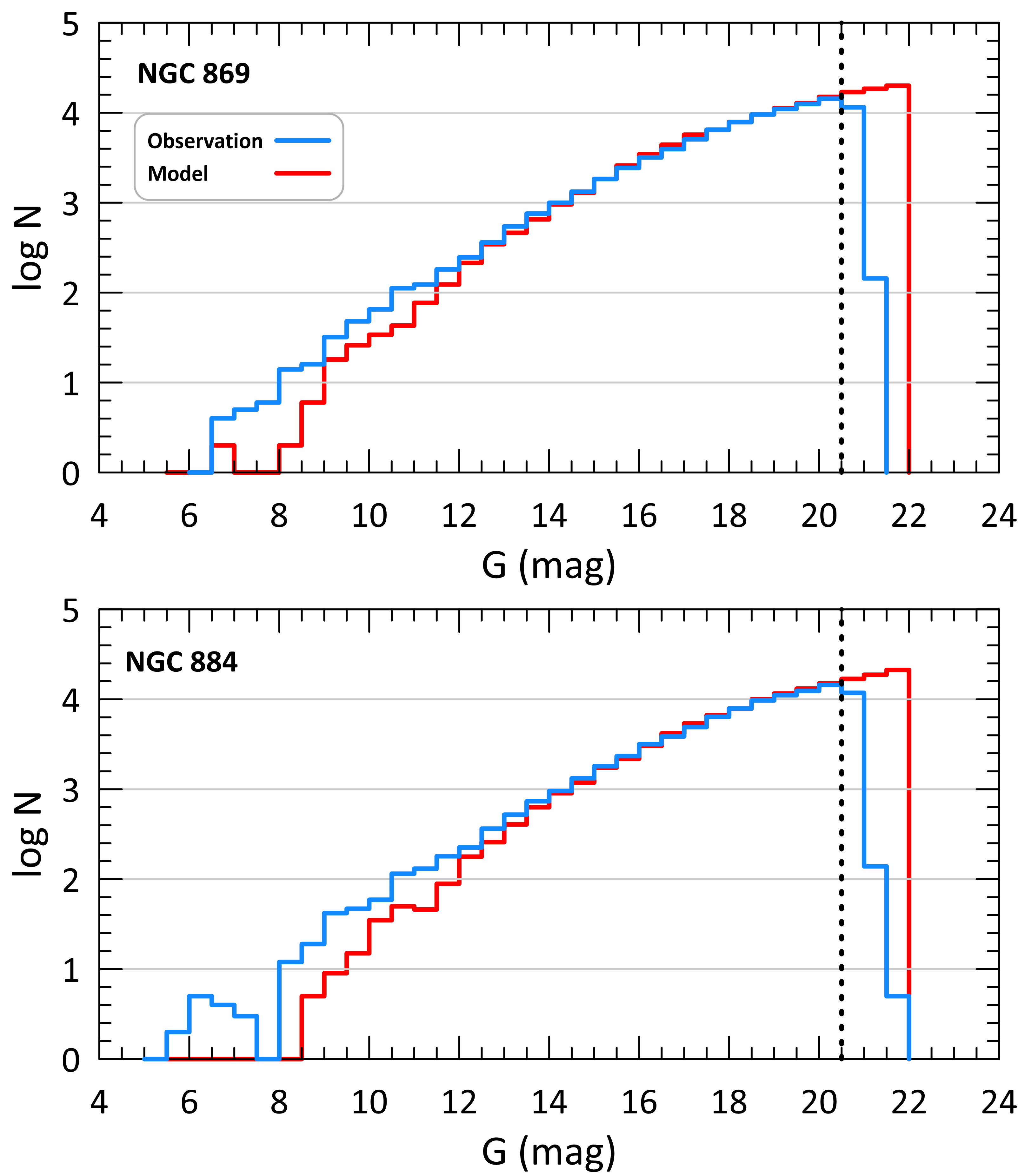}
\caption{The $G-$band star count histograms for NGC\,869 and NGC\,884 are presented, where the observational data are shown in black and the model predictions in red; the black dashed lines denote the photometric completeness limits.}
\label{fig:completness_limit}
\end{figure}

\section{Data Analyses}\label{data_analyzis}

\subsection{Photometric Completeness Limit}

Accurate determination of structural and astrophysical parameters for both OCs requires establishing their photometric completeness limits. This procedure is vital for minimising potential biases in subsequent analyses. For each OC, the stellar number distributions were examined across different \textit{Gaia} $G$-band apparent magnitude intervals. As shown in Figure~\ref{fig:completness_limit}, star counts increase with fainter magnitudes up to a peak near $G= 20.5$~mag. Beyond that magnitude limit, the number of detected sources declines, indicating the point at which the data becomes incomplete. 

To assess the photometric completeness levels, synthetic stellar samples were generated for the fields of both OCs using the Besançon Galaxy Model\footnote{\url{https://model.obs-besancon.fr/}} \citep{Robin03}, which incorporates the three-dimensional extinction map of \citet{Marshall06} and thus provides realistic extinction conditions along the relevant lines of sight. The simulations, carried out over the magnitude range $5 < G~{\rm (mag)} \leq 22$, were designed to reproduce the distribution of the observed star counts and employed solid angles of 0.445 deg$^2$ for each cluster region. Comparison of the synthetic and observed star distributions (Figure~\ref{fig:completness_limit}) shows that the excess of observed stars at bright $G$ magnitudes represents the magnitude domain in which cluster stars are fully sampled, whereas magnitudes at which the model begins to overpredict the observations define the photometric completeness limits. This evaluation confirms that the present data exhibit no loss of completeness, consistent with previous findings. Therefore, the completeness $G-$band apparent magnitudes of both OCs examined in this study were adopted as 20.5 mag.

To estimate photometric uncertainties, the errors provided in the {\it Gaia} DR3 catalogue were considered as interval-based uncertainties. Within defined $G$-band apparent magnitude bins, the mean $G$-band apparent magnitudes and corresponding $G_{\rm BP}-G_{\rm RP}$ colour indices of stars located in the OCs region were computed. At the photometric completeness limit of $G = 20.5$ mag, the mean internal uncertainty in the $G$-band apparent magnitude was determined to be 0.007 mag, while the mean uncertainty in the $G_{\rm BP} - G_{\rm RP}$ colour index was found to be 0.107 mag. A detailed overview of the mean photometric errors across the $G$ magnitude range for both OCs is listed in Table~\ref{tab:photometric_errors}.

These calculations are particularly important as the \textit{Gaia} EDR3/DR3 photometric uncertainties represent a general calibration model based on an all-sky survey \citep{Riello2021}, which may not fully reflect local observing conditions such as crowding and background contamination in dense cluster regions. An inspection of our internal errors reveals that in the apparent magnitude range of $G < 18$, the mean uncertainty is $\sigma_{\rm G} \approx 0.003$\,mag, which remains close to the nominal precision limits presented by \citet{Riello2021}. At the faint end ($G>20$ mag), however, errors increase more rapidly ($\sigma_{\rm G} \approx 0.009 - 0.027$\,mag) compared to the all-sky average due to the expected crowding effect. Notably, the uncertainty in the $G_{\rm BP} - G_{\rm RP}$ colour index reaches $\sim 0.2$\,mag at the limit of $G>20$ mag, which is consistent with the trends in \citet{Riello2021} and justifies the careful treatment of this limit for isochrone fitting.

\begin{table}
\centering
\setlength{\tabcolsep}{16pt}
  \centering
  \caption{Photometric uncertainties in $G$-band apparent magnitudes and $G_{\rm BP} - G_{\rm RP}$ colour indices for stars located in the cluster field.}
    \begin{tabular}{cccc}
      \hline
  $G$ & $N$ & $\sigma_{\rm G}$ & $\sigma_{G_{\rm BP}-G_{\rm RP}}$\\
(mag) &  & (mag) & (mag) \\
\hline
(~5, 12]  & ~~~~~392& 0.003 & 0.006 \\ 
(12, 14]  & ~~1,393 & 0.003 & 0.005 \\ 
(14, 15]  & ~~1,788 & 0.003 & 0.005 \\ 
(15, 16]  & ~~3,420 & 0.003 & 0.006 \\ 
(16, 17]  & ~~5,954 & 0.003 & 0.010 \\ 
(17, 18]  & ~~9,666 & 0.003 & 0.017 \\ 
(18, 19]  & 15,536  & 0.003 & 0.038 \\ 
(19, 20]  & 21,579  & 0.004 & 0.080 \\ 
(20, 21]  & 28,442  & 0.009 & 0.206 \\ 
(21, 22]  & ~~3,869 & 0.027 & 0.446 \\ 
   \hline
    \end{tabular}%
  \label{tab:photometric_errors}%
\end{table}%

\subsection{Structural Parameters}\label{section_structural}

To characterise the spatial distribution and structural properties of NGC\,869 and NGC\,884, a radial density profile (RDP) analysis was performed. The area surrounding each OC was partitioned into a series of concentric annuli, with the OC center as the origin. For each annulus, the surface stellar density $\rho(r)$ was computed using the relation $R_i = N_i / A_i$, where $N_i$ is the number of stars contained within the $i^\mathrm{th}$ ring and $A_i$ is the corresponding area. All annuli were defined such that their outer radii satisfy $r\leq 40'$, ensuring that each ring lies entirely within the selected region and has a complete, well-defined area. Only stars brighter than the photometric completeness limit of $G=20.5$~mag were included in the calculation to avoid biases due to incompleteness.

\begin{figure}
\centering
\includegraphics[width=0.7\linewidth]{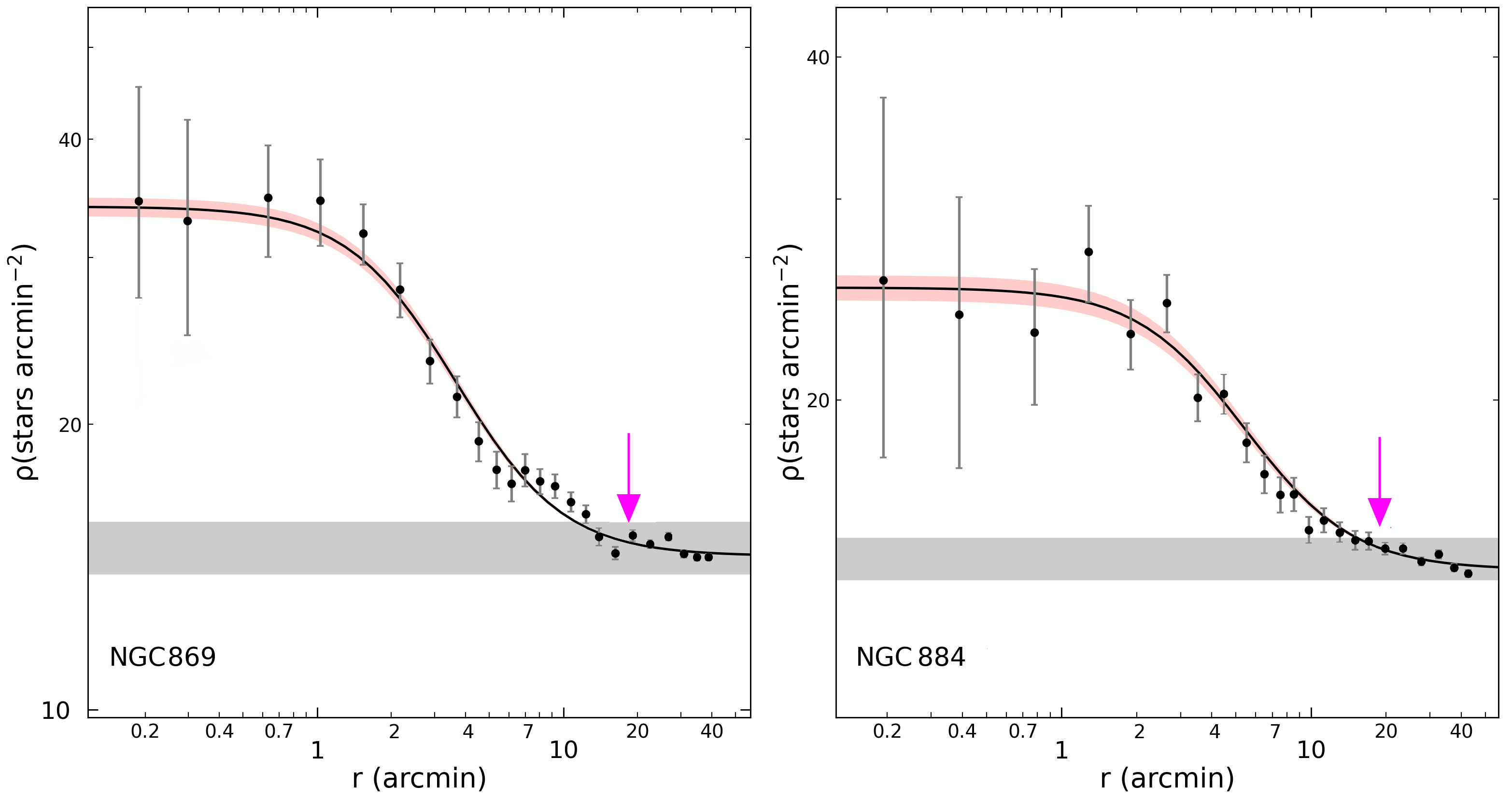}
\caption{RDPs for NGC\,869 (left panel) and NGC\,884 (right panel) with best-fit King models (black lines). The magenta arrows indicate the limiting radii; the gray bands show background density levels. The $1\sigma$ fitting uncertainty is shown as a red-shaded band around the King profile.}
\label{rdp}
\end{figure}

The radial distribution of stellar density was derived by counting stars at successive angular distances from the cluster center, aiming to determine the structural properties of the OCs in this study. The King profile was adopted for the RDP analysis, as it is widely used in the literature owing to its empirical success in describing the stellar density distribution from the cluster center to its outskirts \citep{king62}, and its effectiveness in deriving key structural parameters such as the core and limiting radii \citep{Bonatto2005}. The observed RDPs were compared with \citet{king62} and the structural parameters of the OCs were determined. The King profile model is described as follows:
\begin{equation}
\rho(r)=f_{\rm bg}+\frac{f_0}{1+(r/r_{\rm c})^2},
\end{equation}
where $r$ denotes the radial distance, $r_{\rm c}$ is the core radius, $f_{\rm 0}$ is the central stellar density, and $f_{\rm bg}$ is the background field star density. A chi-squared minimization technique was employed to identify the optimal model parameters, with the final estimates corresponding to the values that minimize the $\chi^2$. An examination of the RDP yielded an observed limiting radius, $r_{\rm lim}^{\rm obs}$, of approximately $20^{\prime}$. The estimated structural parameters obtained from the King model fitting for each OC are illustrated in Figure \ref{rdp}, and the results are given in Table \ref{tab:rdp}.
\begin{equation}
r_{\rm lim} = r_{\rm c} \sqrt{\frac{f_{\rm 0}}{3\sigma_{\rm bg}} - 1}.
\end{equation}

Applying this relation with our fitted parameters resulted in a theoretical limiting radius of $r_{\rm lim}$=19.15$\pm$0.84 and $r_{\rm lim}$=19.52$\pm$0.32 arcmin for NGC\,869 and NGC\,884, respectively. The close agreement between this theoretical estimate and the observed value supports the internal consistency and robustness of the derived structural parameters. When Figure~\ref{rdp} is investigated carefully, a bump in the stellar number density can be seen in the RDP of NGC 869 at a distance of 7-10 arcminutes from the OC center. This raises the question of whether stars belonging to the NGC 884 OC are mixing into NGC\,869. The analyses related to this issue will become clearer after the cluster member stars are identified in Section~\ref{sec:membership} of this study.

In the structural characterisation of young and dense OCs such as NGC\,869 and NGC\,884 the density contrast parameter plays an important role in quantifying the degree of stellar concentration against the background field. Following the formulation introduced by \cite{Bica2005}, the density-contrast parameter is defined as $\delta_{\rm c} = 1 + f_{\rm 0}/f_{\rm bg}$, where $f_{\rm 0}$ and $f_{\rm bg}$ represent the central and background stellar densities, respectively. In this study, the density-contrast parameters derived for NGC\,869 ($\delta_{\rm c}=2.34\pm 0.02$) and NGC\,884 ($\delta_{\rm c}=1.77\pm0.05$) indicate relatively moderate central stellar concentrations in both OCs. For comparison, several Galactic OCs analysed in previous studies exhibit similarly low to moderate density-contrast values. For example, the density-contrast parameter calculated for NGC\,188, which exhibits a relatively high central concentration, is $\delta_{\rm c}=21.5$ \citep{Bonatto2005}. In contrast, \citet{Bica2005}, who analysed five OCs located in the third Galactic quadrant, reported density-contrast values ranging from $\delta_{\rm c}=6.4$ (Czernik\,31) to $\delta_{\rm c}=11.2$ (Trumpler\,13), highlighting that low to intermediate density-contrast values can also be observed among OCs. Similarly, \citet{Camargo2010} demonstrated that young and moderately populated OCs commonly present $\delta_{\rm c} \lesssim 3$, reflecting relatively diffuse core structures. In this context, the density-contrast parameters obtained for NGC\,869 and NGC\,884 are consistent with those reported for other Galactic OCs. This suggests that both OCs may represent dynamically young systems or possess less centrally concentrated core structures compared to more compact stellar systems.

The concentration parameter ($C=r_{\rm tidal}/r_{\rm c}$) is an indicator of a cluster’s internal structure and dynamical state, influenced by internal evolution and the Galactic environment. Although the classical definition uses the tidal radius ($r_{\rm tidal}$) \citep{king1966}, many studies instead adopt the limiting radius ($r_{\rm lim}$) due to the difficulty of measuring $r_{\rm tidal}$, a practical approach also used by \citet{Hill06}, \citet{Munoz10}, and \citet{Miocchi13}. For NGC\,869 and NGC\,884, we obtain concentration parameters of $6.49\pm1.13$ and $4.14\pm0.44$, respectively. These values are higher than the lower concentration limit typically found in dynamically young and weakly concentrated OCs, for which $C \simeq 2$-$3$ is commonly reported in the literature \citep[e.g.,][]{Piskunov2007, Kharchenko2013}. However, they remain below the concentrations characteristic of dynamically evolved systems, indicating that both OCs exhibit moderate central concentration consistent with their young ages and early-stage dynamical evolution. All structural parameters are also provided in Table~\ref{tab:rdp}.

\begin{center}
\begin{table}
\setlength{\tabcolsep}{5pt}
\renewcommand{\arraystretch}{1}
\centering
\caption{Structural parameters of NGC\,869 and NGC\,884 estimated from King model fitting.}
\begin{tabular}{lccc}
\toprule
Parameter & NGC\,869 & NGC\,884 \\
\hline
$r_{\rm c}$ (arcmin) &~~2.95 $\pm$ 0.26 &~~4.71 $\pm$ 0.60 \\
$r_{\rm c}$ (pc) &~~2.04 $\pm$ 0.18 & ~~3.11 $\pm$ 0.40\\
$r_{\rm lim}$ (arcmin) & 19.15 $\pm$ 0.84 & 19.52 $\pm$ 0.32 \\
$r_{\rm lim}$ (pc) & 13.24 $\pm$ 0.59 & 12.90 $\pm$ 0.21 \\
$f_{\rm 0}$ (stars arcmin$^{-2}$) & 19.43 $\pm$ 0.26 & 10.91 $\pm$ 0.64 \\
$f_{\rm bg}$ (stars arcmin$^{-2}$) & 14.52 $\pm$ 0.15 & 14.18 $\pm$ 0.20 \\
$\delta_{\rm c}$  & $2.34\pm 0.02$ & $1.77 \pm 0.05$ \\
$C$ &$6.49\pm1.13$  & $4.14\pm0.44$\\
\hline
\end{tabular}
\label{tab:rdp}
\end{table}
\end{center}

\subsection{Membership Determination}
\label{sec:membership}
A reliable characterisation of the astrophysical parameters of NGC\,869 and NGC\,884 requires an accurate identification of their stellar members, particularly given their location in the dense stellar fields of the Galactic plane, where field-star contamination is significant. Since genuine OC members are coeval and share common kinematic properties, such as proper motions and parallaxes, these features provide a robust basis for distinguishing them from foreground and background stars. To identify probable cluster members, we employed the Unsupervised Photometric Membership Assignment in Stellar Clusters (UPMASK) algorithm \citep{Krone-Martins_2014}. This method is designed to determine cluster membership probabilities by combining positional and photometric information without requiring prior assumptions about the underlying distributions. {\sc UPMASK} operates through an iterative process that utilizes $k$-means clustering and statistical resampling to assess the spatial coherence of stellar groups in multi-dimensional parameter space. It is particularly effective for OCs observed by {\it Gaia}, as it allows for robust member selection even in crowded fields and in the presence of field star contamination.

In this study, we used the {\sc UPMASK} methodology to perform a detailed membership analysis for both OCs. The analysis employed astrometric data from the {\it Gaia} DR3 catalogue, including equatorial coordinates ($\alpha, \delta$), proper-motion components ($\mu_{\alpha} \cos \delta$, $\mu_{\delta}$), and trigonometric parallaxes ($\varpi$), together with their associated uncertainties. {\sc UPMASK} identifies statistically significant clustering of stars consistent with the expected distribution of cluster members by applying the $k$-means clustering algorithm in a multidimensional parameter space. The optimal number of $k$-means clusters was determined by examining the stability of the resulting membership probabilities for different $k$ values. Cluster memberships corresponding to the value at which probabilities no longer changed significantly with increasing $k$ were adopted. Based on this criterion, we selected $k=27$ for NGC\,869 and $k=28$ for NGC\,884, ensuring transparency and robustness in the $k$-means configuration. Due to its effectiveness, {\sc UPMASK} is widely employed in the literature for determining cluster membership \citep[e.g.,][]{Cantat-Gaudin_2018, Castro-Ginard_2018, Cantat_2020, Akbulut_2021, Yontan_2021, Yontan_2022, Wang_2022, Yontan_tek_2023, Karagoz2025}.

\begin{figure}
\centering
\includegraphics[width=0.55\linewidth]{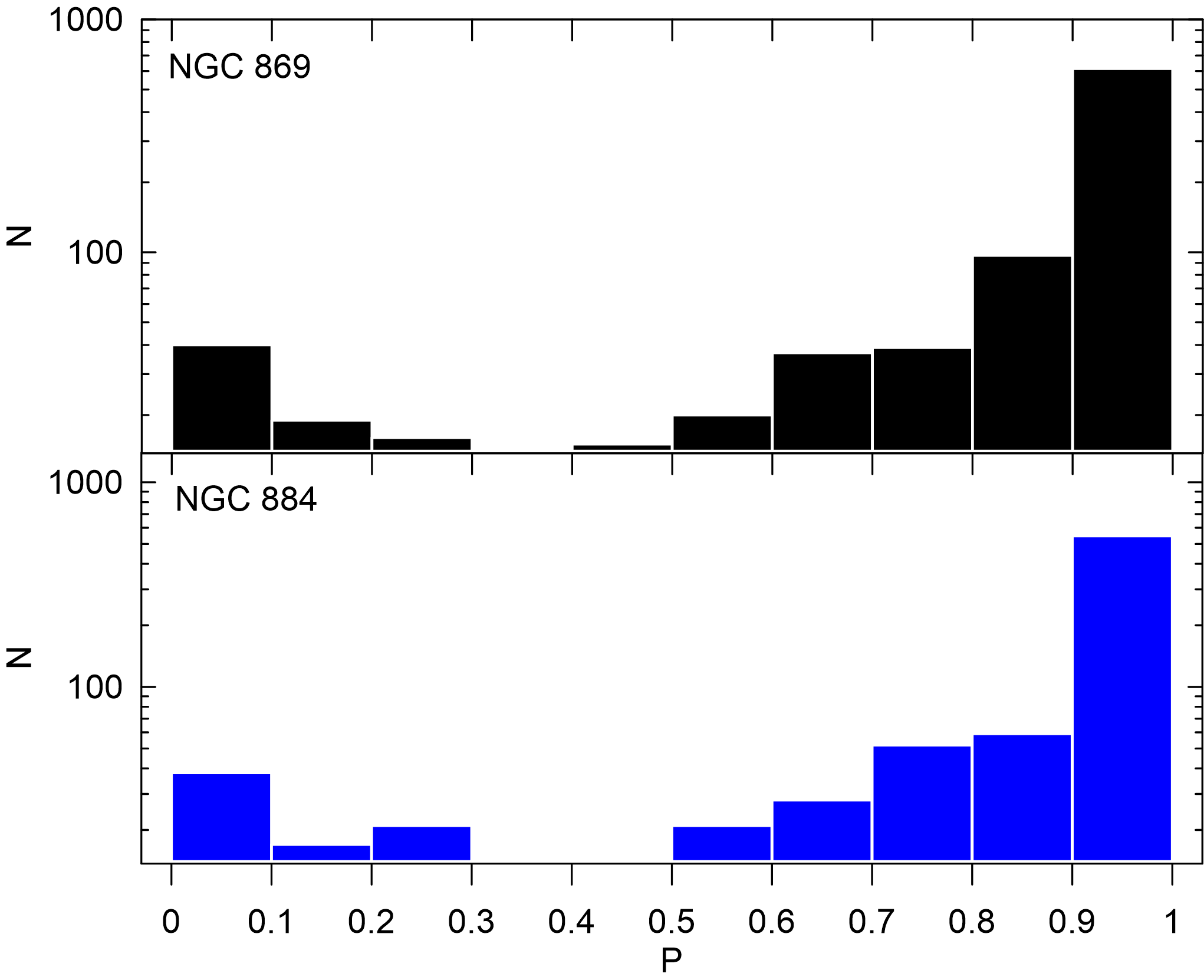}
\caption{Distribution of membership probabilities assigned by the {\sc UPMASK} algorithm for stars in the directions of the NGC\,869 and NGC\,884.}
\label{fig:prob}
\end{figure}

A total of 92,038 stars within the defined search radius around both OCs were initially considered. After filtering based on astrometric precision and photometric completeness (limiting magnitude $G \leq 20.5$ mag), the {\sc UPMASK} algorithm was executed, assigning membership probabilities in the range $0 \leq P \leq 1$ to the remaining stars. The membership probability distributions for both OCs are shown in Figure~\ref{fig:prob}, revealing a clear excess of high-probability stars compared to the field-star baseline, with the majority of stars exhibiting probabilities greater than $P=0.5$. Based on these probabilities, 808 and 707 stars were classified as probable members ($P \geq 0.5$) of NGC\,869 and NGC\,884, respectively.
Only stars satisfying the criterion $P \geq 0.5$ were retained for all subsequent analyses presented in the following sections of this study.

\begin{figure}
\centering
\includegraphics[width=0.45\linewidth]{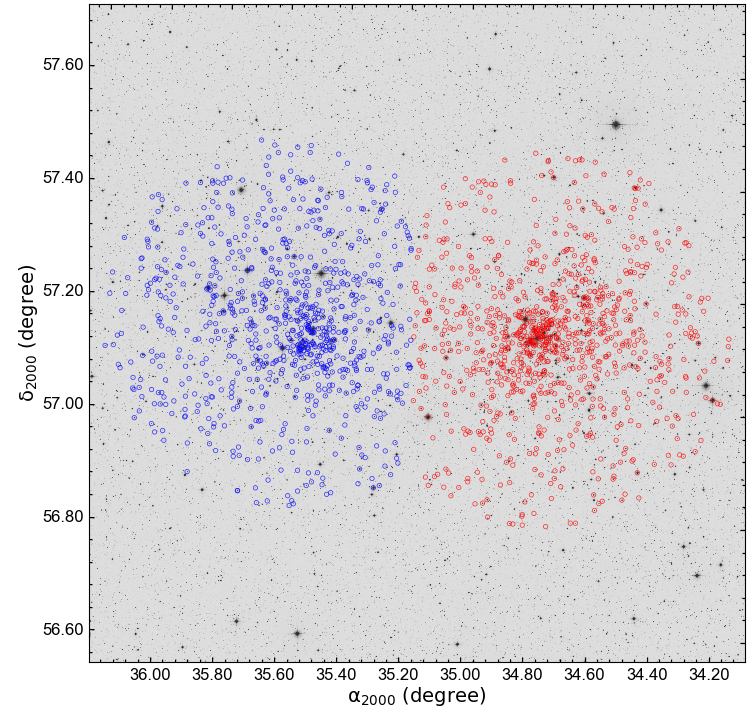}
\caption{Sky chart showing the spatial distribution of members for NGC 869 (black) and NGC 884 (red). The background image is taken from the STScI DSS.}
\label{fig:members_radec}
\end{figure}

Figure~\ref{fig:members_radec} shows the sky positions of stars with membership probabilities $P\geq 0.5$ for both OCs. The stars tested for dual membership exhibit almost no spatial overlap, indicating that there is no population contamination between the OCs. This finding demonstrates that the bump-like feature observed in Figure~\ref{rdp} at a distance of 7-10 arcmin from the center of the NGC\,869 is not caused by contamination from member stars of NGC\,884. Instead, it originates from the spatial distribution of field stars along the line of sight toward the OC.

\begin{figure*}[h]
\centering
\includegraphics[width=.7\linewidth]{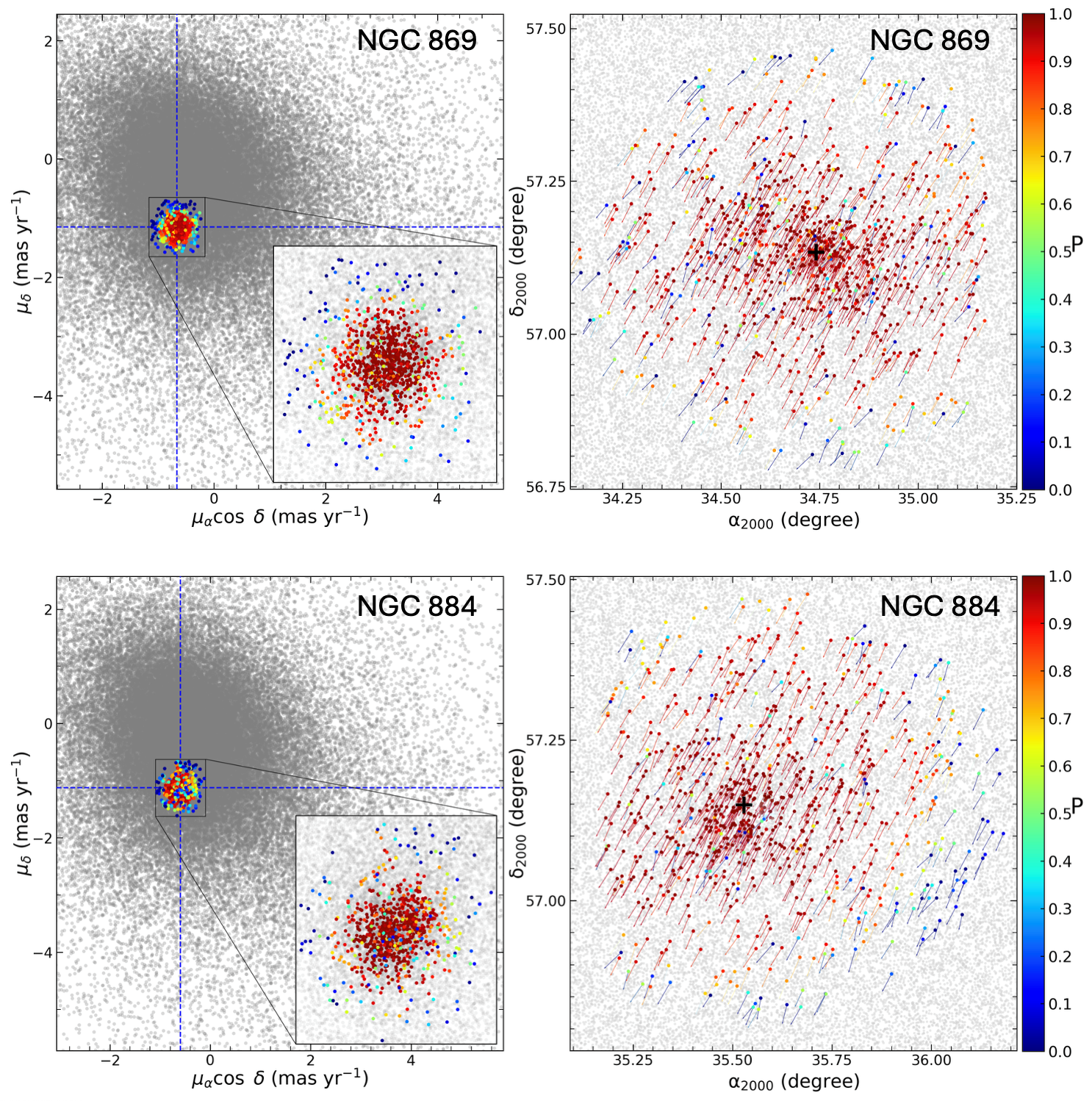}
\caption{The panels show the VPDs and moving vectors of the OC members on the equatorial-coordinate plane, respectively, for the NGC\,869 (upper panels) and NGC\,884 (lower panels). Colour-scaled circles and vectors indicate the likely cluster members and their corresponding movement directions. The zoomed-in boxes in the VPDs indicate the distribution of the OCs according to field stars (gray circles). The intersection of black dashed lines (left panels) indicates the mean proper motion values, whereas the black plus symbols represent the center of the OCs on the equatorial coordinate planes.}
\label{fig:vpd}
\end{figure*}

\subsection{Astrometric Parameters}\label{sec:astrometric}

To further investigate the kinematic coherence of the identified members, we constructed vector point diagrams (VPDs) for both OCs, shown in Figure~\ref{fig:vpd}. These diagrams illustrate the proper-motion components ($\mu_{\alpha} \cos \delta, \mu_{\delta}$) distribution of stars in the VPD. In both cases, the OCs manifest as compact overdensities that are distinguishable from the surrounding star field population. The mean proper-motion components were estimated using stars with $P\geq 0.5$, yielding ($\mu_{\alpha} \cos \delta, \mu_{\delta}$) = (-0.662 $\pm$ 0.022, -1.150 $\pm$ 0.027) mas yr$^{-1}$ for NGC\,869 and (-0.593 $\pm$ 0.025, -1.128 $\pm$ 0.028) mas yr$^{-1}$ for NGC\,884. These results are consistent with previous findings in the literature \citep[c.f.][]{Cantat_2020, Hunt_24}, thereby reinforcing the reliability of the membership determination process. Furthermore, we analysed the {\it Gaia}-based trigonometric parallax values of the confirmed member stars in NGC\,869 and NGC\,884.

\begin{figure}
\centering
\includegraphics[width=0.6\linewidth]{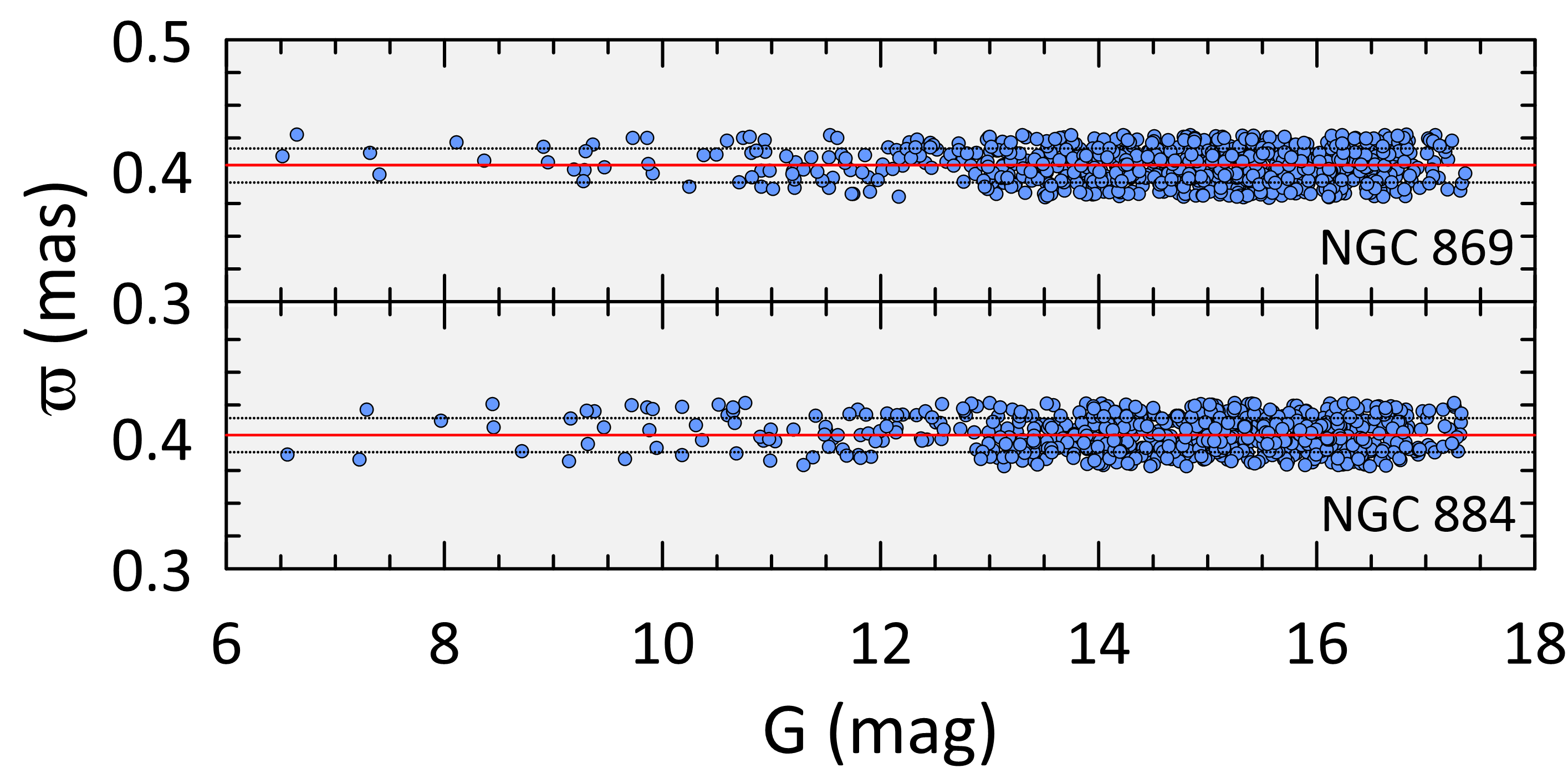}
\caption{Positions of OC member stars with $\sigma_{\varpi}/\varpi\leq 0.2$ in the $\varpi \times G$ diagrams. The red line indicates the median trigonometric parallax of the stars, while the black dashed lines represent the $\pm 1\sigma$ deviation from the median values.}
\label{fig:plx}
\end{figure}

The mean trigonometric parallaxes of both OCs were determined by analysing the distribution of cluster member stars with relative parallax error $\sigma_{\varpi}/\varpi\leq 0.2$ in the trigonometric parallax versus magnitude ($\varpi \times G$) diagram in Figure~\ref{fig:plx}. Under the applied constraints, 728 and 624 stars were identified as likely members of NGC 869 and NGC 884, respectively. These stars have reliably measured trigonometric parallaxes and are detected down to a limiting magnitude of $G=17.5$. As can be seen from Figure~\ref{fig:plx}, the cluster members span a wide range of magnitudes while being concentrated around similar $\varpi$ values, supporting their membership in the respective cluster. The analysis indicates that the median trigonometric parallaxes of the NGC\,869 and NGC\,884, respectively, are 0.404 $\pm$ 0.013 and 0.402 $\pm$ 0.013 mas. To convert these parallax values into heliocentric distances, we used the relation $d({\rm pc}) = 1000 / \varpi~({\rm mas})$. Accordingly, the distances were computed as $d_{\varpi}$ = 2475 $\pm$ 80 pc for NGC \,869 and $d_{\varpi}$ = 2488 $\pm$ 80 pc for NGC\,884.

In this study, the trigonometric parallax values calculated for the NGC\,869 and NGC\,884 were compared with those reported in previous studies based on {\it Gaia} DR2 and {\it Gaia} DR3 astrometric data. \citet{Cantat_2020} reported trigonometric parallax values of $0.399 \pm0.042$ mas and $0.398 \pm 0.038$ mas, while \cite{Dias_2021} provided estimates of $0.400 \pm0.043$ mas and $0.398 \pm 0.039$ mas for the NGC\,869 and NGC\,884, respectively. In this study, the use of {\it Gaia} DR3 data enabled the estimation of trigonometric parallax values with uncertainties approximately three times smaller than early \textit{Gaia} DR2 studies, and also our findings are in agreement with recent studies such as \citet{Hunt_24} reported values of $0.404 \pm 0.019$ mas and $0.402 \pm 0.019$ mas.

\subsection{Astrophysical Parameters Estimation} \label{sec:astrophysical}
To estimate the fundamental parameters of both OCs, we employed two independent and complementary approaches. The first is a Bayesian framework based on Markov Chain Monte Carlo (\texttt{MCMC}) sampling, applied to \textit{Gaia}-based colour-magnitude diagrams (CMDs). This method allows the simultaneous exploration of the parameter space, including reddening, distance modulus, metallicity, and age, while fully accounting for observational uncertainties and parameter degeneracies. The posterior distributions obtained through \texttt{MCMC} provide statistically rigorous constraints on the clusters’ astrophysical properties. The second method utilises spectral energy distribution (SED) fitting, incorporating multi-band photometry to model the observed fluxes of cluster member stars. By comparing the results obtained from both techniques, we aim to achieve a more robust and consistent characterization of the OCs' astrophysical properties.

\subsubsection{MCMC Simulations}
The estimation of the fundamental parameters for NGC\,869 and NGC\,884 was carried out through the construction and analysis of CMDs, which are powerful diagnostics in OC studies \citep[see also,][]{Bilir_2006, Bilir_2010, Bilir_2016, Bostanci_2015, Bostanci_2018, Yontan_2015, Yontan_2019, Ak_2016, Banks_2020, Koc_2022}. These diagrams clearly display evolutionary sequences, such as the main sequence (MS) and turn-off point, enabling precise determination of cluster parameters.

\begin{figure*}
\centering
\includegraphics[width=0.95\linewidth]{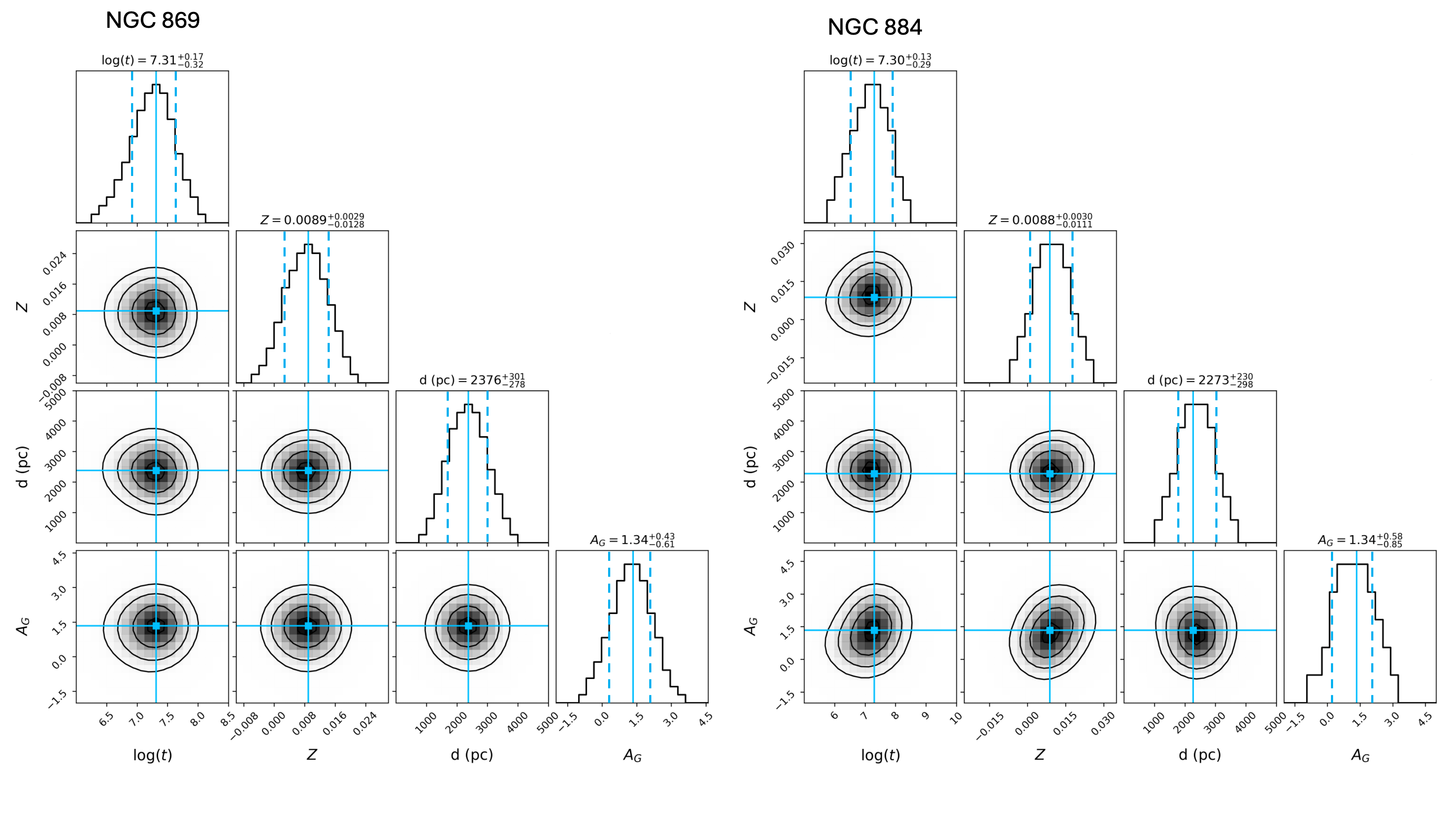}
\caption{Corner diagram illustrating the posterior distributions of the inferred astrophysical parameters for NGC 869 and NGC 884 derived from the \texttt{MCMC} analysis. The solid black lines mark the median values of the parameters, whereas the dashed black lines indicate the corresponding $16^{\rm th}$ and $84^{\rm th}$ percentile confidence intervals.}
\label{fig:mcmc}
\end{figure*}

Precise constraints on key astrophysical quantities such as cluster age, chemical composition, distance modulus, and line-of-sight extinction are crucial for interpreting the formation pathways and evolutionary stages of OCs. To obtain these parameters in a self-consistent way, we applied a Bayesian framework based on a \texttt{MCMC} technique, following the methodology outlined by \citet{Tanik2025}, coupled with the \texttt{PARSEC} stellar evolutionary models \citep{Bressan_2012}. Within this framework, the \texttt{MCMC} sampler explores the joint posterior distribution of the model parameters under the assumption of Gaussian observational uncertainties. The logarithmic likelihood function, adapted from \citet{Tanik2025}, is expressed in Equation~\eqref{equ:log_likelihood}:
\begin{equation}\label{equ:log_likelihood}
\ln \mathcal{L}(\boldsymbol{\theta}) =
\sum_{i=1}^{N_{\mathrm{stars}}} \sum_{X}
\left[
\frac{(m_{X,i} - \hat{m}{X,i})^2}{2\sigma{X,i}^2}
\ln(\sqrt{2\pi},\sigma_{X,i})
\right],
\end{equation}
where $m_{X,i}$ denotes the observed magnitudes in the $G$, $G_{\rm BP}$, and $G_{\rm RP}$ passbands, $\hat{m}{X,i}$ corresponds to the model predictions from the stellar grid, $\sigma{X,i}$ represents the photometric uncertainties, and $\boldsymbol{\theta}$ contains the parameters of interest (log(age), $A_{\rm G}$, distance, and metallicity). The analysis incorporates the high-probability members of each cluster, ensuring that the posterior distribution reflects the intrinsic cluster properties as closely as possible. To efficiently sample this posterior, we employed the \texttt{emcee} ensemble sampler \citep{Foreman-Mackey2013}, assigning one walker per most-probable member star. Each chain was evolved over 5,000 iterations, ensuring adequate convergence. The resulting posterior distributions were used to derive the best-fitting parameter estimates and their associated 1$\sigma$ uncertainties in Figure~\ref{fig:mcmc}, and the fundamental parameters calculated for the OCs are listed in Table~\ref{tab:Final_table}.

\begin{figure*}
\centering
\includegraphics[width=0.95\linewidth]{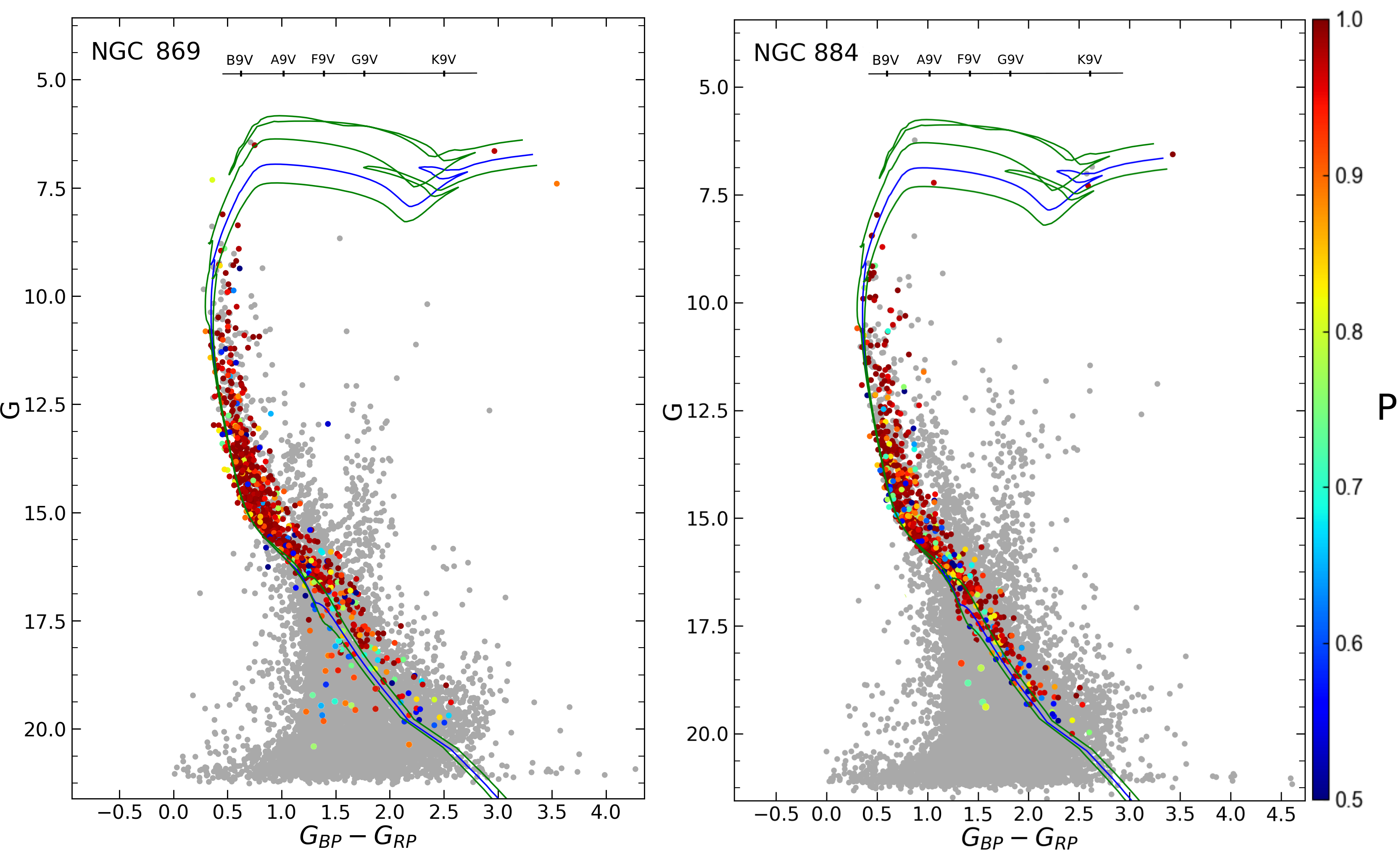}
\caption{CMDs of NGC\,869 and NGC\,884 showing OC members. Stars with $0.5\leq P \leq 1$ are shown in different colour codes, and field stars are displayed in grey circles. The best-fitting {\sc PARSEC} isochrones (black) and their associated uncertainty ranges (green) correspond to the ages derived for both OCs. Spectral type classes of the stars are shown along the top axis of the CMDs, following the classification scheme of \citet{Pecaut2013}.}
\label{fig:CMD}
\end{figure*}

Using the fundamental parameters such as $G$-band extinction, distance, metallicity, and age obtained via the \texttt{MCMC} method, theoretical isochrones from the {\sc PARSEC} database \citep{Bressan_2012} were fitted to the observed CMDs of both OCs. Stars with high-cluster membership probabilities and field stars along the line of sight to the OCs were marked on the $G \times (G_{\rm BP} - G_{\rm RP})$ CMDs constructed from {\it Gaia} photometry, and the corresponding {\sc PARSEC} isochrones based on the parameters derived using the \texttt{MCMC} method were overplotted in Figure~\ref{fig:CMD}. In addition, spectral types of stars, adapted schematically from \citet{Pecaut2013}, are also shown on the CMDs. An examination of the positions of cluster member stars fitted with {\sc PARSEC} isochrones reveals that both OCs contain early-type stars. Additionally, the presence of pre-main-sequence stars, which have not yet reached the main sequence, is noteworthy. This finding provides supporting evidence for the young ages of the OCs. Furthermore, considering the spectral type classifications given by \citet{Pecaut2013}, it has been determined that the stars approaching the main sequence are approximately of early F-spectral type.

\begin{table*}
	\renewcommand{\arraystretch}{0.8}
	\setlength{\tabcolsep}{14pt}
	\centering
	\caption{The fundamental parameters determined for both OCs in this study via \texttt{MCMC} and SED methods.}
	{\normalsize
		\begin{tabular}{lcc}
\toprule
Parameter & NGC\,869 & NGC\,884\\
\hline
\multicolumn{3}{c}{Astrometric Parameters}\\
\hline
($\alpha,~\delta)_{\rm J2000}$                          & 02:18:57.84, $+$57:08:02.40 & 02:22:20.16, $+$57:08:56.40    \\
			($l, b)_{\rm J2000}$ ($^{\rm o}$)           & 134.6257, -3.7372           & 135.0552, -3.5683              \\    
            Number of members                           &  808                        &  707                           \\
			$\mu_{\alpha}\cos \delta$ (mas yr$^{-1}$)   & -$0.662  \pm 0.022$         &-$0.593  \pm 0.025$             \\
			$\mu_{\delta}$ (mas yr$^{-1}$)              & -$1.150  \pm 0.027$         & -$1.128  \pm 0.028$            \\
      $\varpi$ (mas) $(\sigma_\varpi/\varpi \leq 0.2)$  & $0.404 \pm 0.013$ ($N$=728) & $0.402 \pm 0.013$ ($N$=624)    \\
			$d_{\varpi}$ (pc)                           & $2475\pm 80$                & $2488\pm 80$                   \\
            $V_{\rm R}$ (km s$^{-1}$)                   & -$38.17 \pm 3.23$ ($N$=13)  & -$40.21\pm 2.95$ ($N$=14)      \\
\hline
\multicolumn{3}{c}{Astrophysics Parameters - \texttt{MCMC} method}\\
\hline
            Number of members                           &  808                          &  707                         \\
    $E(B-V)$ (mag)                                      & $0.516 ^{+0.17}_{-0.24}$      & $0.516 ^{+0.22}_{-0.33}$     \\
    $E(G_{\rm BP}-G_{\rm RP})$ (mag)                    & $0.72^{+0.23}_{-0.33}$        & $0.72^{+0.31}_{-0.46}$       \\
    $A_{\rm G}$ (mag)                                   & $1.34^{+0.43}_{-0.61}$        & $1.34^{+0.58}_{-0.85}$       \\
    ${\rm [Fe/H]}$ (dex)                                & $-0.24\pm 0.12$               & $-0.25\pm 0.12$              \\
                $Z$                                     & $0.0089^{+0.0029}_{-0.0128}$  & $0.0088^{+0.0031}_{-0.0111}$ \\
    $(G-M_{\rm G})$ (mag)                               & $13.22 ^{+0.68}_{-0.88}$      & $13.12 ^{+0.79}_{-1.15}$     \\
            $d_{\rm iso}$ (pc)                          & $2376^{+301}_{-278}$          & $2273^{+230}_{-298}$         \\
                $\log$~($t/\rm yr$)                     & $7.31^{+0.17}_{-0.32}$        & $7.30^{+0.13}_{-0.29}$       \\
\hline
\multicolumn{3}{c}{Astrophysics Parameters - SED method}\\
\hline
Number of stars                                         & 349               & 286                 \\
            $E(B-V)$ (mag)                              & $0.484\pm0.032$   &  $0.481\pm0.032 $   \\
			$A_{\rm V}$ (mag)                           & $1.50\pm 0.10$    & $1.49\pm 0.10$      \\
			$[{\rm Fe/H}]$ (dex)                        & -$0.25\pm 0.03$   & -$0.25\pm 0.03$     \\
			$(V-M_{\rm V})$ (mag)                       & $13.381\pm 0.123$ & $13.279\pm 0.129$   \\
			$d_{\rm SED}$ (pc)                          & $2378\pm 25$      &  $2279\pm 31$       \\
                $Z$                                     & $0.0088\pm0.0009$ & $0.0088\pm0.0009$   \\
			 $\log$~($t/\rm yr$)                           & $7.30 \pm 0.09$   &  $7.30 \pm 0.09$    \\
\hline
		\end{tabular}%
	} 
	\label{tab:Final_table}%
\end{table*}%

From the colour excesses, the corresponding extinction values in the $G$ band were computed using $A_{\rm G} = 1.8626 \times E(G_{\rm BP}-G_{\rm RP})$. These {\it Gaia}-based colour excesses and extinction estimates were subsequently transformed into the {\it UBV} photometric system through the relations $E(G_{\rm BP}-G_{\rm RP}) = 1.41 \times E(B-V)$ and $A_{\rm G} = 0.83626 \times A_{\rm V}$ \citep{Canbay_2023}. The apparent distance moduli obtained from the posterior distributions correspond to \texttt{MCMC}-inferred distances of $2376^{+301}_{-278}$~pc for NGC\,869 and $2273^{+230}_{-298}$~pc for NGC\,884.

The relation provided by Bovy\footnote{\url{https://github.com/jobovy/isodist/blob/master/isodist/Isochrone.py}}was used to convert the heavy-element abundances of the OCs into iron abundances ([Fe/H]), and it is calibrated for \texttt{PARSEC} stellar isochrones \citep{Bressan_2012}. This relation has been successfully applied in several previous studies \citep[e.g.,][]{gokmen_2023, Yontan_2023a, Yontan_2023b, Cakmak_2024}. In this approach, an intermediate variable $Z_{\rm x}$ is first computed as:
\begin{equation}
Z_{\rm x} = \frac{{\rm Z}}{0.7515 - 2.78 \times {\rm Z}}
\label{equ: Zx}
\end{equation}
followed by the calculation of the OC iron abundances:
\begin{equation}
{\rm [Fe/H]} = \log \left({\rm Z_{\rm x}} \right) - \log \left( \frac{Z_{\odot}}{1 - 0.248 - 2.78 \times Z_{\odot}} \right)
\label{equ: [Fe/H]}
\end{equation}
In these expressions, $Z_\odot=0.0152$ denotes the solar metallicity. The heavy-element abundances estimated using the \texttt{MCMC} method were converted into iron abundances via Equations~\ref{equ: Zx} and \ref{equ: [Fe/H]}, yielding metallicities ([Fe/H]) of $-0.24\pm 0.12$ and $-0.25\pm 0.12$ dex for NGC\,869 and NGC\,884, respectively.

For the NGC\,869 and NGC\,884, the colour excesses determined in the {\it Gaia} photometric system were transformed into the {\it UBV} system using the relations provided by \citet{Canbay_2023}, yielding $E(B-V)=0.516 ^{+0.17}_{-0.24}$ and $E(B-V)=0.516 ^{+0.22}_{-0.33}$ mag, respectively. These results are in agreement with those reported in the literature. In particular, \citet{Dias_2021} derived values of $E(B-V) = 0.564 \pm 0.012$ mag for NGC\,869 and $0.551 \pm 0.006$ mag for NGC\,884, which are consistent with our findings within the quoted uncertainties. Moreover, \citet{Zhong_2020} reported colour excess of $E(B-V)=0.52$ mag for both OCs, further supporting the reliability of our estimates. 

In this study, it is found that the distances calculated using the trigonometric parallax method for both OCs differ by 99-215 pc compared to those calculated using the \texttt{MCMC} method (see Table~\ref{tab:Final_table}). This discrepancy becomes particularly significant for astronomical objects located beyond 2 kpc and is mainly caused by biases in the stellar trigonometric parallax measurements. As demonstrated by \citet{Plevne2020}, this effect can be clearly seen by comparing {\it Gaia} trigonometric parallax data with the bias-corrected distances of \citet{Bailer-Jones2018} at distances of approximately 2 kpc from the Sun. In this study, the distances of both OCs derived from trigonometric parallaxes ($\sim$ 2.5 kpc) are measured to be, on average, $\cong$ 200 pc closer than those obtained using a Bayesian approach. This explains the differences between the distances derived using the two different methods in this study. For the reasons stated above, the distances calculated using the \texttt{MCMC} method are adopted in this study. The distances estimated for both OCs in this study show general consistency with values reported in recent literature. For instance, \citet{Hunt_24} reported distances of $2279 \pm 6$ pc and $2293 \pm 7$ pc for NGC\,869 and NGC\,884, respectively. Similarly, \citet{Dias_2021} obtained distances of $2246 \pm 64$ pc and $2150 \pm 103$ pc for NGC\,869 and NGC\,884, respectively, which are consistent within the uncertainties. In addition, \citet{Zhong_2020} estimated distances of 2300 pc for NGC\,869 and 2200 pc for NGC\,884, providing further support to our derived values.

The metallicity adopted for both OCs in this study, [Fe/H]$\cong -0.25\pm 0.12$ dex, is consistent with the limited spectroscopic constraints available for NGC\,869 and NGC\,884. Although high-resolution spectroscopic analyses for those clusters are scarce, large-scale spectroscopic surveys such as LAMOST provide statistically robust metallicity estimates. For example, \citet{Zhong_2020} reported mildly subsolar metallicities for both clusters based on LAMOST DR5 spectra, supporting our adopted value. While these survey-based measurements are not high-resolution abundance analyses, they offer the most direct observational constraints currently available, and our metallicity assumption aligns well with these results.

The estimated logarithmic ages of $\log(t/{\rm yr}) \cong 7.30 \pm 0.15$ for both OCs are also consistent with literature values. For example, \citet{Dias_2021} reported $\log(t/{\rm yr}) = 7.11 \pm 0.04$ for NGC\,869 and $7.19 \pm 0.05$ for NGC\,884, while other studies \citep[e.g.][]{Cantat_2020, Song2022} suggest values within a similar range, corroborating the young age nature of both OCs. Additionally, \citet{Zhong_2020} estimated an age of approximately 14 Myr for NGC\,869 and 15 Myr for NGC\,884, which are within the uncertainty limits of our results and confirm the coeval nature of both OCs.

\subsubsection{Bayesian SED Modeling of OC Member Stars}

A detailed SED analysis was conducted for the young both OCs using the {\sc ARIADNE} Python package, a robust tool designed to perform Bayesian inference on stellar photometric data. This framework incorporates a suite of theoretical stellar atmosphere models, spanning a wide range of effective temperatures, surface gravities, and metallicities, which have been pre-convolved with an extensive set of standard photometric filter transmission curves \citep{Vines2022}. The sampling of the posterior probability distributions, as well as the estimation of the Bayesian evidence for competing models, was achieved through the implementation of the nested sampling algorithm provided by the {\sc dynesty} library \citep{Higson2019, Speagle2020}. 

To build the broadband SEDs, we compiled all available photometry from the optical to the infrared regime and performed a careful cross-matching across surveys. We adopted the $Gaia$~DR3 $G$, $G_{\rm BP}$, and $G_{\rm RP}$ measurements \citep{Gaia_DR3} as the astrometric reference frame and matched them to 2MASS near-infrared data ($J$, $H$, $K_{\rm s}$; \citealt{Skrutskie06}) using a $1^{\prime\prime}$ radius. Mid-infrared fluxes from {\it WISE} ($W1$-$W4$; \citealt{Wright2010}) were added through a positional cross-match with typical separations $<0^{\prime\prime}.5$, ensuring that only sources with clean photometric quality flags were retained. When available, Pan-STARRS1 and APASS measurements were incorporated to extend the spectral coverage and to reduce degeneracies in temperature and extinction.

The fitting procedure was executed in static mode, ensuring efficient exploration of the multi-dimensional parameter space without dynamic adjustment of the number of live points. We adopted a configuration of 500 live points, with the convergence criterion set to a log-evidence difference of 0.5 between successive iterations. To enhance the sampling efficiency and accuracy of the posterior distributions, we employed the random-walk sampling strategy alongside a multi-ellipsoidal bounding method. Two complementary stellar atmosphere model grids, PHOENIX \citep{Husser2013} and Castelli-Kurucz \citep{Castelli_2003}, were utilized to reduce systematic dependencies associated with any single model set. For extinction correction, the \citet{Fitzpatrick_1999} reddening law was adopted, which provides reliable performance across optical and near-infrared bands. Each SED fit was computed using 100,000 posterior samples, enabling robust statistical characterisation of the model parameters.

For the SED fitting of individual cluster members, we adopted the geometric distance estimates provided by the Bayesian Bailer-Jones inference framework \citep{BailerJones2021}, which are automatically retrieved and implemented by the {\sc ARIADNE} SED-fitting tool \citep{Vines2022}. These distances offer a significantly improved treatment of parallax uncertainties at $\sim2$ kpc compared to direct inversion of parallaxes and therefore provide stable distance priors without fixing all stars to the same cluster value. Line-of-sight extinction values were taken from the \citet{Schlafly2011} dust maps. Using these priors, the SED fitting constrains the key stellar parameters, effective temperature ($T_{\rm eff}$), surface gravity ($\log g$), metallicity ([Fe/H]), interstellar extinction ($A_{\rm V}$), and stellar radius ($R$), within a statistically consistent Bayesian framework appropriate for heterogeneous stellar populations in young OCs \citep{Alzahrani2025b, Bisht2025}.

\begin{figure*}
\centering
\includegraphics[width=0.9\linewidth]{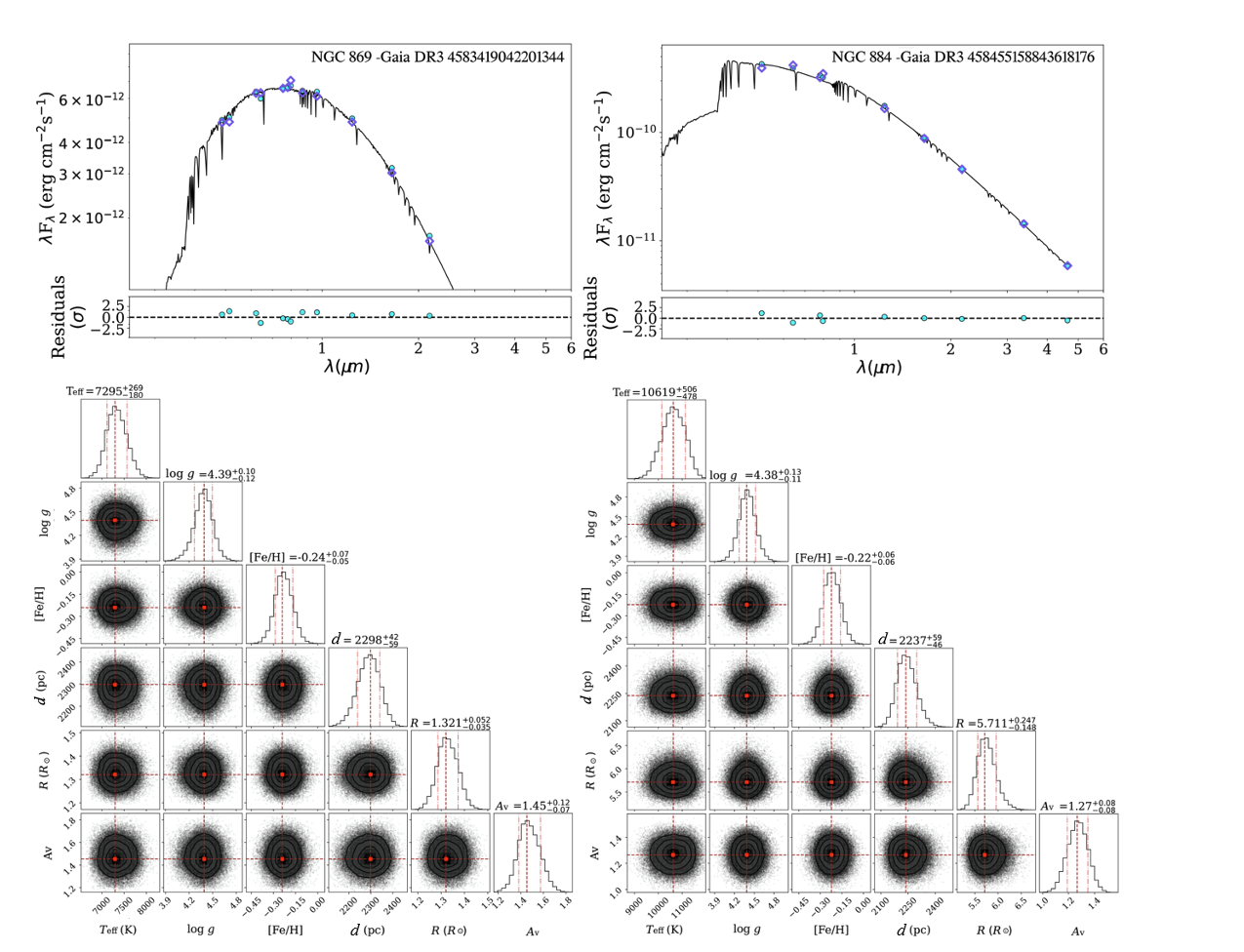}
\caption{The upper panels present the SED fits for representative selected one-member stars from each OC, while the bottom panels display the corresponding corner histograms and distributions of the best-fit astrophysical parameters.}
\label{Fig:SED_corners}
\end{figure*}

Astrometric reliability was assessed using the \texttt{RUWE} parameter, and stars with $\texttt{RUWE} > 1.4$ were excluded from the analysis. In addition to the \texttt{RUWE} criterion, we implemented two complementary filters from the \textit{Gaia} DR3 database: the “variable” flag was used to discard sources with significant photometric variability that could compromise the stability of the SED fit, and the “duplicated source” flag was employed to eliminate entries affected by crowding or catalogue-level duplication. Following the application of these quality control measures, 259 stars were excluded from NGC\,869 and 251 from NGC\,884. Of the remaining stars, 349 in NGC\,869 and 286 in NGC\,884 were found to have a sufficient number of photometric measurements with reliable magnitudes to allow for the construction of robust SEDs. Consequently, the SED analyses were limited to these subsets. As representative examples, one member star from each cluster was selected, and their corresponding SED fits together with the posterior distributions of the fitted parameters are shown in the form of SED distributions and corner plots in Figure~\ref{Fig:SED_corners}.

\begin{figure*}
\centering
\includegraphics[width=0.9\linewidth]{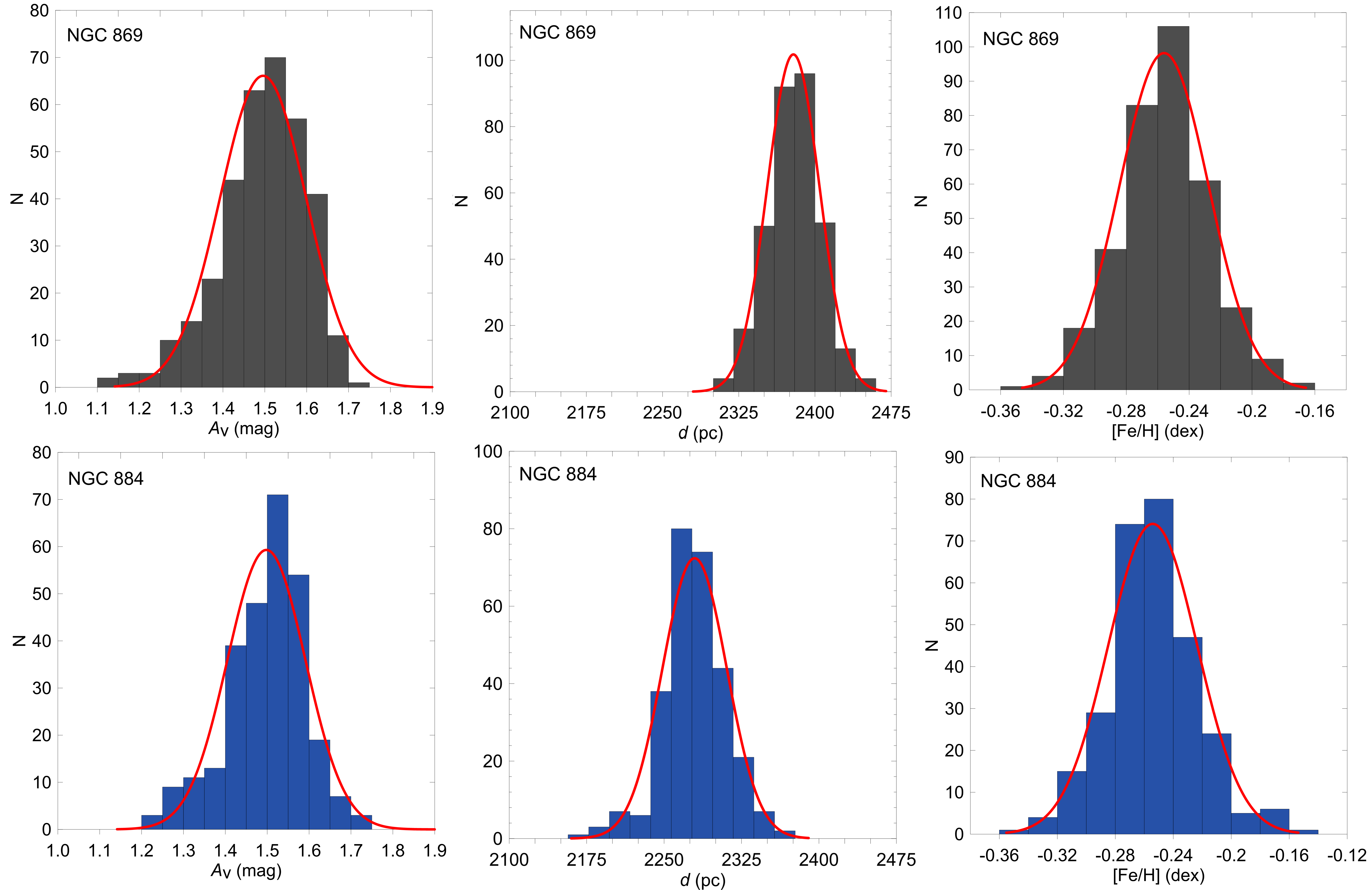}
\caption{Histograms illustrating the distributions of $V$-band extinction ($A_{\rm v}$), and heliocentric distance ($d$) and metallicity ([Fe/H]) for stellar members of the NGC\,869 and NGC\,884, as derived from SED fitting. The red dashed lines correspond to standard Gaussian profiles.}
\label{Fig:SED_histogram}
\end{figure*}

The fundamental parameters derived from the SED analysis for the member stars of both OCs are presented in Table~\ref{tab:Final_table} and shown in Figure~\ref{Fig:SED_histogram}. As with many young OCs, the most significant sources of uncertainty are reddening and distance estimates. The colour excess ($E(B-V)$) values obtained from the SED analyses for the member stars of NGC\,869 and NGC\,884 were adapted to the {\it Gaia} photometric system using appropriate conversion coefficients. The resulting extinction, distance, and metallicity distributions exhibit narrow spreads, indicating a relatively homogeneous interstellar medium across both OCs. The mean extinction values are found to be $A_{\rm V} = 1.50 \pm 0.10$~mag for NGC~869 and $A_{\rm V} = 1.49 \pm 0.10$~mag for NGC~884, corresponding to colour excesses of $E(B-V) = 0.484 \pm 0.032$~mag and $0.481 \pm 0.032$~mag, respectively, assuming a standard total-to-selective extinction ratio of $R_{\rm V} = 3.1$ \citep{Cardelli1989}. Both OCs exhibit slightly subsolar metallicities, with a mean value of $[{\rm Fe/H}] = -0.25 \pm 0.03$~dex, corresponding to a metallicity fraction of $Z = 0.0088 \pm 0.0009$. The distance distributions are tightly clustered around mean values of $2378 \pm 25$~pc for NGC~869 and $2279 \pm 31$~pc for NGC~884. These fundamental parameters derived from the SED analysis are fully consistent, within uncertainties, with the values obtained from the \texttt{MCMC}-based isochrone fitting presented in Section~\ref{sec:astrophysical}. The agreement between these two independent approaches demonstrates the robustness of the adopted cluster parameters of both OCs.

\section{Kinematical and Dynamical Orbital analyses}\label{Dynamic_Kinematics}

To estimate the Galactic orbital parameters of both OCs, we utilised the Python library {\sc galpy}\footnote{\url https://galpy.readthedocs.io/en/v1.5.0/}, which operates within the framework of the {\sc MWPotential2014} Galactic potential introduced by \citet{Bovy_2015}. This potential represents the Milky Way as an axisymmetric system composed of three principal components: the bulge, the disc, and the halo. The bulge component follows a spherical power-law density profile as defined by \citet{Bovy_2015}, the disc adopts the axisymmetric formulation described by \citet{Miyamoto_1975}, and the halo structure is modeled after the spherically symmetric dark-matter distribution proposed by \citet{Navarro_1996}. Detailed parameters used for orbital integrations and model fit within the {\sc MWPotential2014} framework are extensively documented in \citet{Bovy_2015}. Based on their findings, we adopt Galactocentric distance of $R_0 = 8.125~\mathrm{kpc}$ \citep{Gravity2018}, and a corresponding solar circular velocity of $V_\odot = 242~\mathrm{km~s^{-1}}$.

To explore the spatial distribution of the OCs within the MW, the Galactocentric distance ($R_\text{gc}$) was computed using the following formula \citep{Guctekin2019}:
\begin{equation}
R_\text{gc} = \sqrt{R_{0}^{2} + (r \cos b)^2 - 2 R_{0} r \cos b \cos l},
\end{equation}
where $R_{0}$ is the distance from the Sun to the Galactic centre, $r$ is the distance to OC, and $l$ and $b$ represent the Galactic longitude and latitude, respectively. The current distances of NGC\,869 and NGC\,884 from the Galactic centre are $9935\pm 237$ and $9862\pm 215$ pc, respectively.

Accurate determination of the space velocities and orbital properties of the OCs under study requires reliable estimates of their mean radial velocities. In this study, we did not restrict our analysis to radial velocity measurements available solely in the {\it Gaia} DR3 catalogue \citep{Gaia_DR3}. Instead, we extended our search by incorporating additional radial velocity measurements from the compilation of \citet{2022AA...659A..95T}, which includes data from large spectroscopic surveys such as LAMOST and APOGEE. Candidate stars were identified by cross-matching the radial velocity catalogues with the OC membership lists and applying strict quality criteria. Specifically, only stars with membership probabilities $P \geq 0.7$ and reliable astrometric solutions characterized by $\texttt{RUWE} \leq 1.4$ were retained. For NGC 869, six stars were initially matched in the catalogue of \citet{2022AA...659A..95T}. However, one star with a low membership probability ($P \simeq 0.1$) and one star flagged as variable were excluded from the analysis. As a result, four stars from this catalogue were retained for NGC 869. For NGC 884, all stars matched in the same catalogue satisfied the membership probability and quality criteria, yielding a final sample of four non-\textit{Gaia} stars. Consequently, a total of eight stars from external spectroscopic surveys (four in each OC) were included in addition to the {\it Gaia} DR3 radial velocity measurements.

In total, the final radial velocity samples comprise 13 stars for NGC 869 and 14 stars for NGC 884, with 9 and 10 stars, respectively, originating from the {\it Gaia} DR3 catalogue, and the remaining four stars in each OC taken from external spectroscopic surveys. Using the weighted mean method described by \citet{Carrera_2022}, we derived mean radial velocities of $V_{\rm R} = -38.17 \pm 3.23$ km s$^{-1}$ for NGC 869 and $V_{\rm R} = -40.21 \pm 2.95$ km s$^{-1}$ for NGC 884. The detailed stellar properties and survey origins are listed in Table~\ref{tab:RV_table}, while the adopted mean values are listed in Table~\ref{tab:Final_table}.

\begin{table}
\centering
\footnotesize
\setlength{\tabcolsep}{2.5pt}
\renewcommand{\arraystretch}{0.8}
\caption{Radial velocity data for stars with high membership probabilities in NGC~869 and NGC~884. The table lists the {\it Gaia} DR3 source IDs, equatorial coordinates, $G$-band magnitudes, radial velocities, \texttt{RUWE} values, membership probabilities, and the survey providing the radial velocity measurements.}
\label{tab:RV_table}
\begin{tabular}{ccccccclc}
\hline
\hline
Order & Gaia Source ID & $\alpha_{\rm J2000}$ & $\delta_{\rm J2000}$ & $G$ & RV & \texttt{RUWE} & $P$ & Survey* \\
      &               & ($^\circ$)          & ($^\circ$)          & (mag) & (km s$^{-1}$) & & & \\
\hline
\hline
\multicolumn{9}{c}{NGC 869} \\
\hline
01 & 458357741598022528 & 35.1208 & 56.9931 &  6.65 & -45.79$\pm$0.40  & 1.01 & 0.95 & Gaia \\
02 & 458373035989882880 & 34.5629 & 57.1711 &  9.91 &  -8.51$\pm$13.06 & 1.00 & 0.90 & Gaia \\
03 & 458373826264109312 & 34.9170 & 57.0920 &  9.86 & -103.15$\pm$11.96& 1.09 & 0.99 & Gaia \\
04 & 458375235013296000 & 34.8081 & 57.1692 &  6.51 & -43.56$\pm$1.72  & 0.98 & 0.97 & Gaia \\
05 & 458379014584350592 & 34.6173 & 57.2084 &  8.11 & -65.51$\pm$8.32  & 1.06 & 0.97 & Gaia \\
06 & 458560159122834816 & 34.4875 & 57.1282 & 11.19 & -100.39$\pm$29.55& 1.01 & 0.99 & Gaia \\
07 & 458561052475969920 & 34.4819 & 57.2188 & 12.33 & -20.68$\pm$36.30 & 1.12 & 0.99 & Gaia \\
08 & 458568164941730048 & 34.4304 & 57.3147 & 12.95 & -62.16$\pm$4.81  & 1.05 & 0.77 & Gaia \\
09 & 458569917288566528 & 34.7220 & 57.4213 &  7.40 & -35.09$\pm$0.46  & 0.96 & 0.78 & Gaia \\
10 & 458364652214200064 & 34.7035 & 56.8883 & 14.82 & -13.32$\pm$2.81  & 0.97 & 0.95 & LAMOST \\
11 & 458569401892222592 & 34.6447 & 57.3879 & 14.00 &  -0.62$\pm$2.68  & 1.01 & 0.88 & LAMOST \\
12 & 458362487550472576 & 35.0999 & 57.0491 & 14.98 &  14.90$\pm$1.33  & 0.99 & 0.92 & LAMOST \\
13 & 458374169861447296 & 34.9010 & 57.1203 & 15.72 & -44.15$\pm$1.90  & 0.95 & 0.98 & LAMOST \\
\hline
\multicolumn{9}{c}{NGC 884} \\
\hline
01 & 458407464445792384 & 35.6012 & 57.1095 &  6.56 & -46.18$\pm$0.37  & 1.06 & 0.99 & Gaia \\
02 & 458414542551436672 & 35.7229 & 57.2452 &  7.22 & -50.22$\pm$0.98  & 1.07 & 0.94 & Gaia \\
03 & 458414989228047744 & 35.8505 & 57.2120 &  7.29 & -45.69$\pm$0.14  & 1.02 & 0.95 & Gaia \\
04 & 458454606008605312 & 35.5024 & 57.1450 &  7.97 & -41.61$\pm$1.66  & 0.98 & 0.99 & Gaia \\
05 & 458310604343959552 & 35.3395 & 56.8985 & 10.76 & -86.96$\pm$15.42 & 1.31 & 0.87 & Gaia \\
06 & 458413511759296896 & 35.8281 & 57.2162 & 10.98 & -29.48$\pm$20.82 & 1.03 & 0.95 & Gaia \\
07 & 458458454299129088 & 35.2199 & 57.2244 & 11.20 & -47.05$\pm$3.02  & 1.10 & 0.98 & Gaia \\
08 & 458452372625503744 & 35.3888 & 57.0324 & 11.38 & -117.26$\pm$27.10& 1.23 & 0.98 & Gaia \\
09 & 458440071838143616 & 35.9335 & 57.3966 & 12.06 & -44.35$\pm$23.43 & 1.22 & 0.84 & Gaia \\
10 & 458461404928737024 & 35.6199 & 57.2662 & 12.33 & -89.02$\pm$35.49 & 1.18 & 0.96 & Gaia \\
11 & 458413339950196736 & 35.7963 & 57.1685 & 12.26 &  -3.72$\pm$0.35  & 0.98 & 0.94 & APOGEE \\
12 & 458408937611303424 & 35.9408 & 57.0878 & 12.64 &  -9.91$\pm$3.06  & 1.17 & 0.73 & LAMOST \\
13 & 458416879013598592 & 35.8433 & 57.3653 & 14.84 & -24.10$\pm$2.77  & 1.01 & 0.94 & LAMOST \\
14 & 458359670051882240 & 35.2682 & 57.0681 & 15.65 & -18.51$\pm$1.68  & 1.06 & 0.98 & LAMOST \\
\hline
\hline
\end{tabular}
\noindent
\\
(*): Survey denotes the source of the radial velocity measurements: {\it Gaia} (Gaia DR3), LAMOST (Large Sky Area Multi-Object Fibre Spectroscopic Telescope), and APOGEE (Apache Point Observatory Galactic Evolution Experiment).
\end{table}

In comparing the mean radial velocity values obtained in this study, the current era of high-precision astrometric data provided by the {\it Gaia} mission has been considered. \citet{Soubiran_2018}, using {\it Gaia} DR2 data \citep{Gaia_DR2}, reported a radial velocity of $V_{\rm R} = -62.82 \pm 16$ km s$^{-1}$ for NGC\,869 based on a member star, and $V_{\rm R} = -43.01 \pm 5.69$ km s$^{-1}$ for NGC\,884 based on two member stars. Similarly, \citet{Tarricq_2021}, also utilising {\it Gaia} DR2, derived a mean radial velocity of $V_{\rm R} = -44.09 \pm 7.52$ km s$^{-1}$ for NGC\,869 from one member, and $V_{\rm R} = -33.66 \pm 4.03$ km s$^{-1}$ for NGC\,884 from five members. More recently, \citet{Hunt_2023}, using {\it Gaia} DR3 data \citep{Gaia_DR3}, calculated a mean radial velocity of $V_{\rm R} = -67.64 \pm 10.42$ km s$^{-1}$ for NGC\,869 based on six member stars, and $V_{\rm R} = -43.90 \pm 1.61$ km s$^{-1}$ for NGC\,884 based on two members. When considered in the context of these recent studies, the mean radial velocity for NGC\,869 derived in the present work is consistent with the result reported by \citet{Tarricq_2021}, while the value obtained for NGC\,884 shows strong agreement with both \citet{Soubiran_2018} and \citet{Hunt_2023}. It should also be noted that the radial velocity estimates presented in this study are based on measurements from a total of 27 stars across both OCs. 

\subsection{Space Velocity Estimation}
To estimate the mean space velocity components for both OCs we collected the equatorial coordinates ($\alpha$, $\delta$) of the OCs, as reported by \citet{Cantat_2020}, the mean proper-motion components ($\mu_{\alpha}\cos\delta$, $\mu_{\delta}$) determined in Section~\ref{sec:astrometric}, the \texttt{MCMC}-based distances ($d_{\rm iso}$) from Section~\ref{sec:astrophysical}, and the radial velocities ($V_{\rm R}$) estimated in this study (see Table~\ref{tab:Final_table}). The space-velocity components of the NGC\,869 and NGC\,884 were computed using the {\sc galpy} package developed by \citet{Bovy_2015}, and are listed Table~\ref {Tab:Dynamics}). In this context, the $U$ component denotes the velocity directed toward the Galactic centre, $V$ represents the velocity in the direction of Galactic rotation, and $W$ indicates the velocity component perpendicular to the Galactic plane, pointing toward the north Galactic pole \citep{Bilir2012, Doner2023}. The uncertainties associated with the space-velocity components were quantified following the method proposed by \citet{Johnson_1987}. To account for the motion relative to the local standard of rest (LSR), we adopted the correction values of $(8.83\pm0.24, 14.19\pm0.34, 6.57\pm0.21)$ km s$^{-1}$, as reported by \citet{Coskunoglu_2011}. After applying these corrections, the LSR-adjusted velocity components were recalculated as:
$(U, V, W)_{\rm LSR} = (39.63 \pm 1.88, -13.13 \pm 2.20, -5.47 \pm 1.63)\ \mathrm{km\ s^{-1}}$ for NGC\,869 and $(37.73 \pm 2.20, -12.02 \pm 2.29, -5.08 \pm 1.41)\ \mathrm{km\ s^{-1}}$ for NGC\,884. In addition, the total space velocity ($S_{\rm LSR}$) to the LSR for each OC was derived from their respective mean total space velocity components \citep{Coskunoglu2012}.
\begin{align}
S_{\rm LSR} = \sqrt{{U_{\rm LSR}}^2 + {V_{\rm LSR}}^2 + {W_{\rm LSR}}^2}.
\end{align}
Based on these values, the total space velocities relative to the LSR were derived as $S_{\rm LSR} = 42.11\pm3.31~$km~s$^{-1}$ for NGC\,869 and $S_{\rm LSR} = 39.92\pm3.47~$km~s$^{-1}$ for NGC\,884. These results are indicative of kinematic properties typical for young thin-disc stellar populations, as discussed by \citet{Leggett_1992, Yontan_2023a}.

\citet{Soubiran_2018} investigated the kinematic properties of the NGC 869 and NGC 884 using astrometric data from {\it Gaia} DR2 \citep{Gaia_DR2}, reporting space velocity components of $(U, V, W) = (46.96\pm11.29, -43.00\pm11.44, -9.71\pm1.05)$ km s$^{-1}$ for NGC 869 and $(32.72\pm4.02, -29.17\pm4.01, -10.81\pm0.37)$ km s$^{-1}$ for NGC\,884. The findings of the present study reveal a notable divergence from the results of \citet{Soubiran_2018} for NGC\,869, while demonstrating a high level of agreement for NGC\,884. The inconsistency in the case of NGC\,869 is most likely due to the methodological constraint in \citet{Soubiran_2018}, wherein the cluster's radial velocity was estimated using data from one member star. Such a limited sample may not reliably represent the systemic motion of the cluster, thereby introducing potential biases into the derived space velocity components.

\begin{table}
\centering
\renewcommand{\arraystretch}{0.9}
\setlength{\tabcolsep}{5pt}
\caption{The spatial, kinematic, Galactic orbital, and dynamical parameters of both OCs are presented.}
\begin{tabular}{l|cc}
\toprule
Parameter & NGC\,869 & NGC\,884 \\
\hline
\multicolumn{3}{c}{Spatial Parameters} \\
\hline
$X_{\odot}$ (pc)  & -1666$\pm$203 & -1606$\pm$186\\
$Y_{\odot}$ (pc)  & 1687$\pm$206  & 1603$\pm$186\\
$Z_{\odot}$ (pc)  & -155$\pm$19   & -141$\pm$17 \\
$R_{\rm gc}$ (pc) & 9935$\pm$237  & 9862$\pm$215\\
\hline
\multicolumn{3}{c}{Kinematical Parameters} \\
\hline
$\overline{U}$ (km s$^{-1}$) & 30.79$\pm$1.86 & 28.90$\pm$2.19 \\
$\overline{V}$ (km s$^{-1}$) & -27.32$\pm$2.17 & -26.21$\pm$2.27 \\
$\overline{W}$ (km s$^{-1}$) & -12.04$\pm$1.61 & -11.65$\pm$1.390 \\
$U_{\rm LSR}$ (km s$^{-1}$) & 39.63$\pm$1.88 & 37.73$\pm$2.20\\
$V_{\rm LSR}$ (km s$^{-1}$) & -13.13$\pm$2.20 & -12.02$\pm$2.29\\
$W_{\rm LSR}$ (km s$^{-1}$) & -5.47$\pm$1.63 & -5.08$\pm$1.41\\
$S_{\rm LSR}$ (km s$^{-1}$) & 42.11$\pm$3.31 & 39.92$\pm$3.47\\
$\overline{V_x}$ (km s$^{-1}$) & -4.15$\pm$0.50 & -3.65$\pm$0.52 \\
$\overline{V_y}$ (km s$^{-1}$) & -12.44$\pm$3.53 & -11.25$\pm$3.35 \\
$\overline{V_z}$ (km s$^{-1}$) & -41.10$\pm$6.41 & -39.29$\pm$6.27 \\
$A_{\rm o}(^{\rm o})$ & -108.43$\pm$0.09 & -107.96$\pm$0.09 \\
$D_{\rm o}(^{\rm o})$ & -72.30$\pm$0.12 & -73.24$\pm$0.12 \\
\hline
\multicolumn{3}{c}{Galactic Orbital Parameters} \\
\hline
$R_{\rm a}$ (pc)      & 9949$\pm$209    & 9926$\pm$216 \\
$R_{\rm p}$ (pc)      & 9384$\pm$473    & 9444$\pm$428 \\
$R_{\rm m}$ (pc)      & 9667$\pm$341    & 9685$\pm$322 \\
$Z_{\rm max}$ (pc)    & 154$\pm$31      & 143$\pm$27 \\
$e$                   & 0.029$\pm$0.015 & 0.025$\pm$0.012 \\
$P_{\rm orb}$ (Myr)   & 250$\pm$10      & 251$\pm$9 \\
$R_{\rm birth}$ (pc)  & 9876$\pm$148    & 9841$\pm$126 \\
\hline
\multicolumn{3}{c}{Dynamical Parameters} \\
\hline
$T_{\rm relax}$ (Myr) & 29 & 40 \\
$r_{\rm t}$ (pc) & 2.04 & 3.26 \\
$\tau_{\rm ev}$ (Myr) & 4000 & 5200 \\
\hline
\end{tabular}
\label{Tab:Dynamics}
\end{table}

The member stars of OCs typically demonstrate coherent proper-motion vectors that, when extended across the celestial sphere, appear to converge toward a specific region termed the convergent point (CP). This geometrical behavior arises from the shared kinematic motion of OC members through space and offers valuable constraints on the systemic dynamics of the cluster \citep{Elsanhoury2018, Elsanhoury2026}. Estimation of CP coordinates enables researchers to probe the internal kinematic alignment and the dynamical cohesion of the cluster system.

\begin{figure}
\centering
\includegraphics[width=0.75\linewidth]{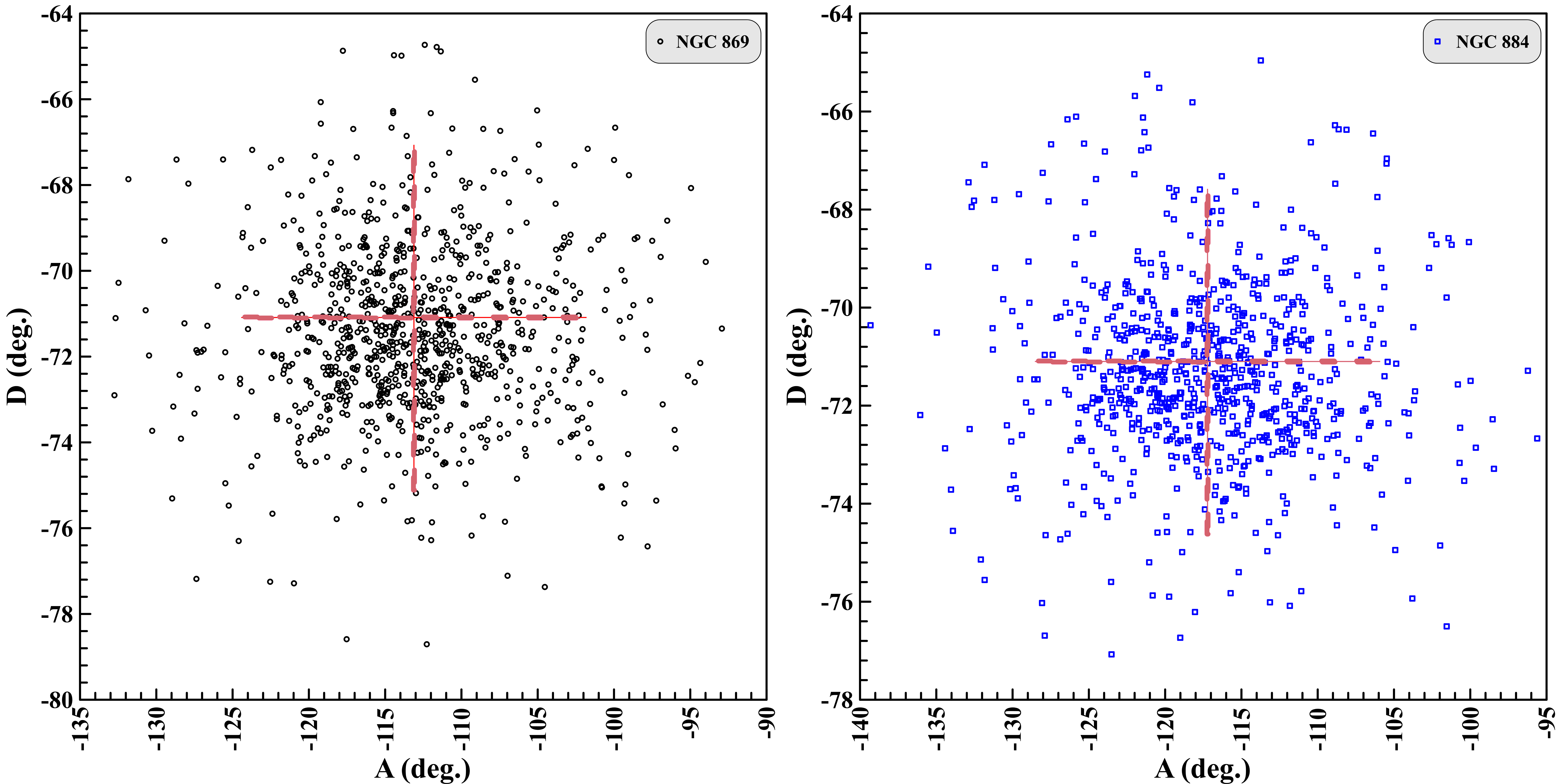}
\caption{CP diagrams illustrating the apex solutions $(A_{\rm o}, D_{\rm o})$ for NGC\,869 and NGC\,884; the apex coordinates are marked by a cross symbol.}
\label{Fig:CP}
\end{figure}

To estimate the CP coordinates for both OCs, we employed the apex-declination diagram (AD-diagram) technique proposed by \citet{Chupina2001, Chupina2006}, a method widely used for stellar systems exhibiting group motion. This approach involves mapping the velocity vector-derived apex coordinates ($A_{\rm o}$, $D_{\rm o}$), where $A_{\rm o}$ and $D_{\rm o}$ indicate the right ascension and declination of the CP, respectively. The intersection region of these vectors for cluster members marks the inferred CP position. In this study, we adopted the analytical framework formulated by \citet{Haroon2025}, which relates the CP coordinates to the mean space velocity components ($\overline{V}_x$, $\overline{V}_y$, $\overline{V}_z$) by trigonometric transformations as follows (see Table~~\ref{Tab:Dynamics}). For each star, the components of the space velocity vector were computed within a right-handed rectangular coordinate system defined with respect to the equatorial reference frame, ensuring consistency with standard astrometric conventions.
\begin{eqnarray}
V_x=-V_{\alpha}\sin\alpha-(V_{\delta} \sin \delta-V_{\rm R} \cos \delta) \cos \alpha, \\ \nonumber
V_y=V_{\alpha} \cos \alpha-(V_{\delta} \sin \delta-V_{\rm R} \cos \delta) \sin \alpha, \\ \nonumber
V_z=V_{\delta} \cos \delta+V_{\rm R} \sin \delta.\\ \nonumber 
\end{eqnarray}
Here, $\alpha$ and $\delta$ are the equatorial coordinates of the cluster member stars, $V_{\rm R}$ is the mean radial velocity of the OCs, and $V_\alpha$ and $V_{\delta}$ denote the components of the total space velocity corresponding to the proper motion in the right ascension ($\mu_{\alpha}\cos \delta$) and declination ($\mu_{\delta}$), respectively, and are defined by the following relations.
\begin{equation}
V_{\alpha}=4.74047\times d_{\rm iso}\times \mu_\alpha \cos \delta, \quad V_\delta=4.74047\times d_{\rm iso}\times \mu_\delta,   
\end{equation}
where $d_{\rm iso}$ is the distance of the cluster calculated by the isochrone fitting method (see Table~\ref{Tab:Dynamics}). The resulting apex positions for NGC\,869 and NGC\,884 are depicted as crosses in Figure~\ref{Fig:CP} with applied the AD-diagram method described in some studies in the literature \citep[e.g.][]{ 2016Ap.....59..246E, 2019SerAJ.198...45E, Elsanhoury2025, Elsanhoury_2024}. Each stellar apex represents the point on the celestial sphere where the spatial velocity vector intersects. The cluster apex coordinates $(A_\mathrm{o}, D_\mathrm{o})$ are computed using:
\begin{align}
A_\mathrm{o} &= \tan^{-1} \left( \frac{\overline{V_y}}{\overline{V_x}} \right), \\
D_\mathrm{o} &= \tan^{-1} \left( \frac{\overline{V_z}}{\sqrt{\overline{V_x}^2 + \overline{V_y}^2}} \right).
\end{align}
Here, following the averaging of the velocity components $\overline{V_x}$, $\overline{V_y}$, and $\overline{V_z}$, the equatorial coordinates of the CP were subsequently determined and listed in Table~\ref{Tab:Dynamics}, and these coordinates were shown in Figure \ref{Fig:CP}.

\subsection{Galactic Orbital Parameters}

To derive the orbital characteristics of both OCs, we used the {\sc MWPotential2014} Galactic potential model. The input parameters included the equatorial coordinates ($\alpha$, $\delta$), the mean proper-motion components ($\mu_{\alpha}\cos\delta$, $\mu_{\delta}$), the distances ($d_{\rm iso}$), and the radial velocities ($V_{\rm R}$) as listed in Table~\ref{tab:Final_table}. To ensure sufficient temporal resolution for determining the minimum separation, orbital integration was performed backward in time using a time step of 0.1 Myr over a total timespan of 2 Gyr. This duration was chosen to ensure complete orbital closure and to obtain robust estimates of the Galactic orbital parameters. The resulting parameters are also presented in Table~\ref{Tab:Dynamics}, where $R_{\rm a}$ and $R_{\rm p}$ correspond to the apogalactic and perigalactic distances, respectively. The orbital eccentricity is denoted by $e$, while $Z_{\rm max}$ indicates the maximum vertical distance from the Galactic plane.

\begin{figure*}
\centering
\includegraphics[width=0.85\linewidth]{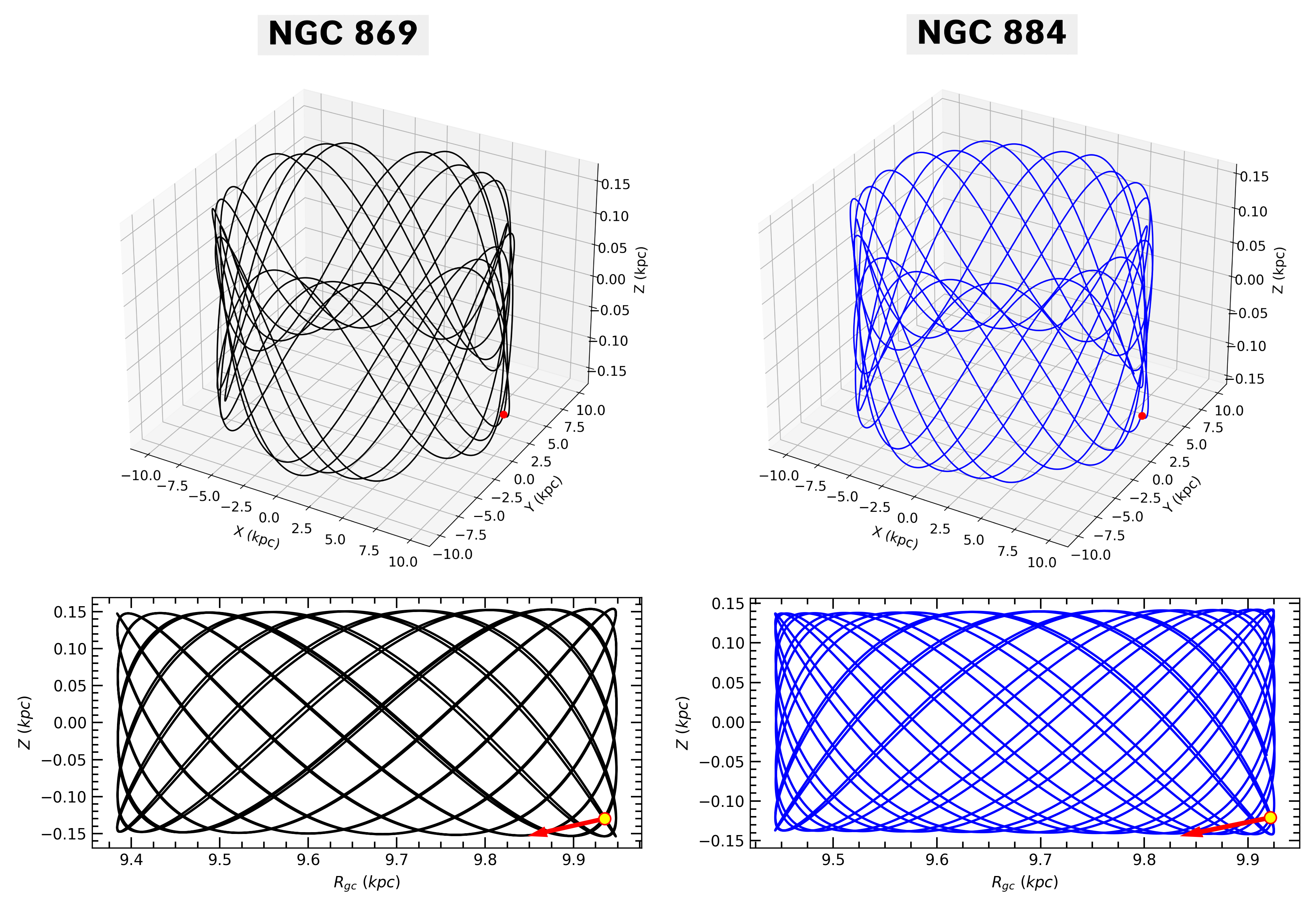}
\caption{The 3D orbits of the OCs around the Galactic centre are shown. In the upper panels, red dashed circles mark the present-day positions of the OCs, while the lower panels display the Galactic orbits of NGC\,869 and NGC\,884 projected onto the $Z \times R_{\rm gc}$ planes. The red arrow indicates the direction of motion for both OCs.}
\label{orbits}
\end{figure*}

While the 3D orbits of the investigated OCs around the Galactic centre, computed using {\sc galpy}, are presented in the upper panels of Figure~\ref{orbits}, the lower panels show the Galactic orbital trajectories of NGC\,884 and NGC\,869, projected onto the $Z \times R_{\rm gc}$ planes. As can be seen from Figure~\ref{orbits}, neither OC deviates significantly from the Galactic plane and follows a similar orbit with a comparable eccentricity around the Galactic centre. As summarised in Table~\ref{Tab:Dynamics}, the orbital analysis shows that NGC\,869 and NGC\,884 follow low-eccentricity, nearly circular orbits near 9-10 kpc, with small vertical excursions ($Z_{\rm max}\sim$140-160 pc) that keep them within the Galactic thin disk. Their similar orbital periods and radial motions further indicate that both clusters share comparable dynamical histories, consistent with formation in a common large-scale Galactic environment. The low eccentricities of both OCs and small deviations from the Galactic plane suggest that NGC\,869 and NGC\,884 are dynamically associated with the thin-disc component of the MW \citep{Tasdemir_2023, Canbay_2025, Cinar_2024, Cinar_2025}.

\begin{figure}
\centering
\includegraphics[width=0.71\linewidth]{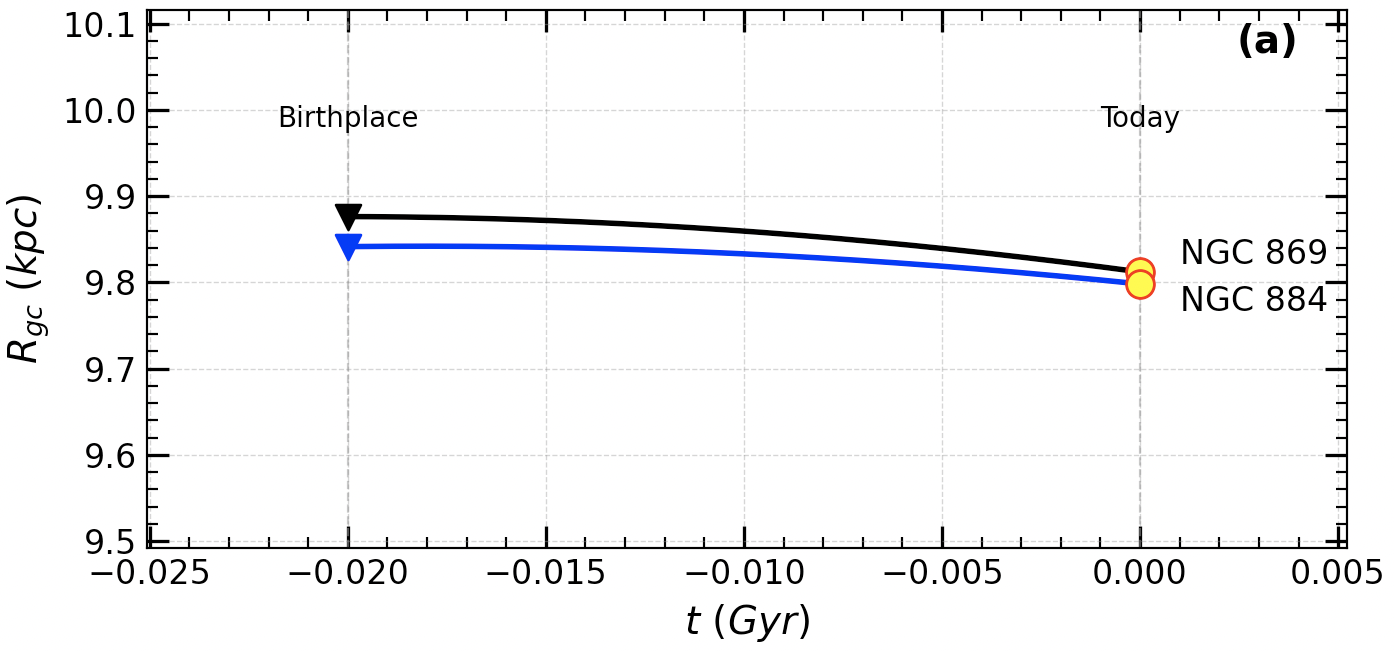}
\includegraphics[width=0.7\linewidth]{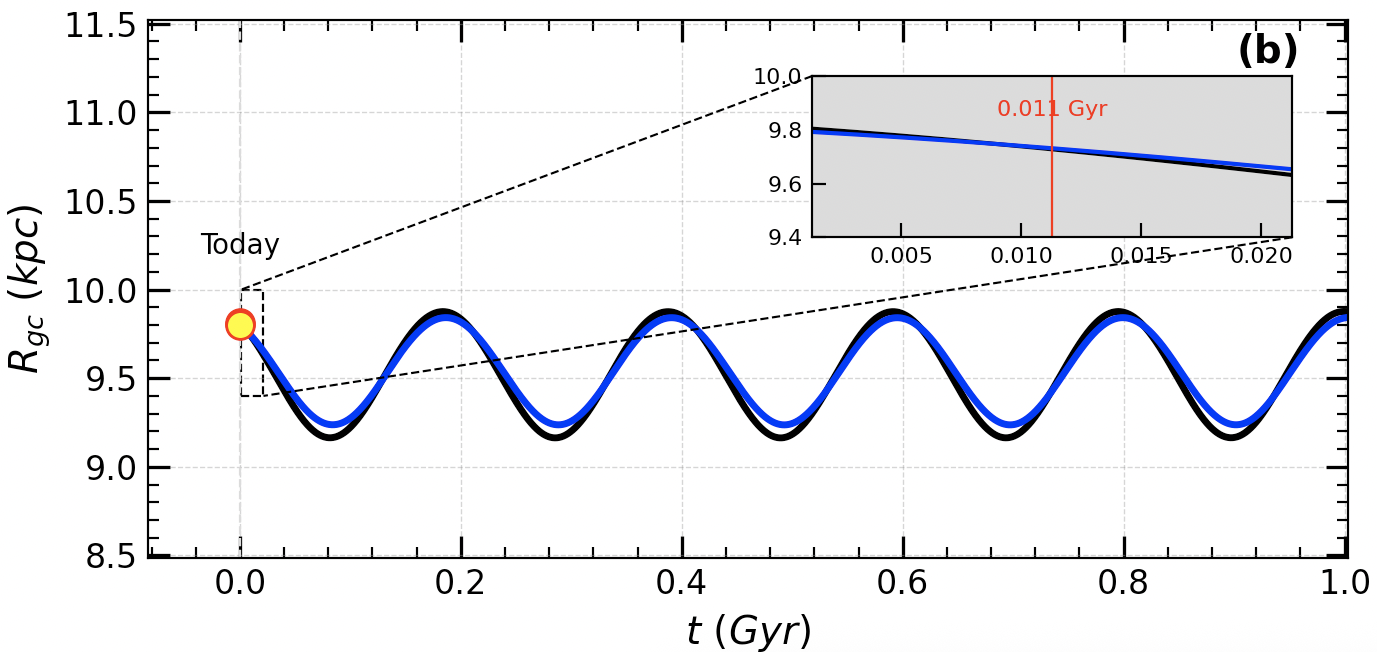}
\includegraphics[width=0.66\linewidth]{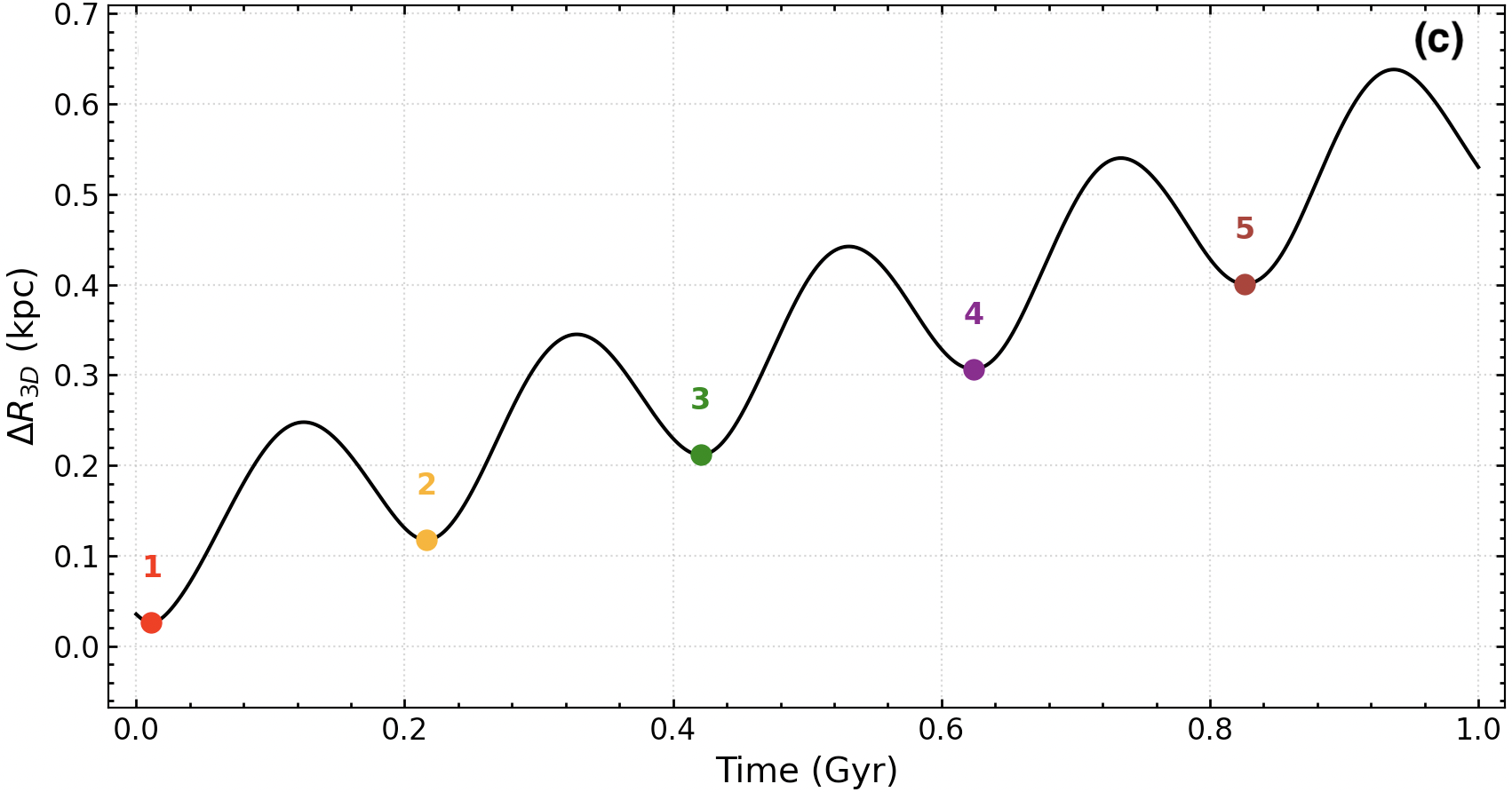}
\caption{The orbital dynamics and separation history of NGC\,869 (black line) and NGC\,884 (black line). 
Panel (a) displays the orbital tracebacks over the past $\sim$20 Myr in the $R_{\rm gc} \times t$ plane, highlighting the estimated birthplace. Filled yellow circles indicate present-day positions ($t=0$), while triangles mark the inferred birthplaces. Panel (b) illustrates the future orbital evolution of both OCs over the next 1 Gyr. The inset provides a zoomed-in view of the pair's closest spatial encounter, which occurred at $t \approx 11$~Myr. Panel (c) shows the temporal evolution of the 3D separation ($\Delta R_{\rm 3D}$) between the two clusters. The numbered colored circles (1--5) correspond to the epochs of local minima (closest approaches) in their mutual distance.}
\label{fig:births}
\end{figure}

To examine whether the two investigated OCs originated from nearby regions of the Galaxy, their present-day positions within the Galaxy were plotted on the $R_{\rm gc} \times t$ plane, and their birthplaces were investigated by tracing their orbits backward in time over a period corresponding to their estimated ages \citep{Yucel_2024, Tasdemir_2025}. In Figure~\ref{fig:births}, the present-day positions of the OCs are indicated by yellow circles, and their estimated birthplaces are marked with triangles, while the shaded regions and dotted lines represent the orbital uncertainties arising from the propagation of errors in the astrometric (proper motions and distances) and spectroscopic (radial velocities) input parameters. Backward orbital integrations under a static Galactic potential were used to estimate the likely birth radii of NGC\,869 and NGC\,884. This classical traceback approach \citep{Akbaba2024} provides a simple means of assessing whether the clusters have experienced significant radial migration. Our calculations indicate that both OCs were probably formed at Galactocentric distances very close to their current locations, with inferred birthplace radii of $R_{\rm birth} = 9876 \pm 148$ and $9841 \pm 126$ pc, respectively.

The similarity in their orbital parameters, along with their proximity in both position and age, further supports the scenario of a common origin. The results of the dynamical orbital integration are consistent with the hypothesis that NGC\,884 and NGC\,869 formed in nearby regions of the Galactic disc and have subsequently evolved as a co-moving cluster pair. Such consistency between past orbital properties and future orbital convergence strengthens the interpretation that these clusters not only share a common origin but may also experience renewed dynamical interaction. Figure~\ref{fig:births}b displays orbital integration of NGC\,869 and NGC\,884, indicating that, although the two OCs do not currently exhibit strong signs of mutual gravitational binding, their orbits will bring them significantly closer in the future. Specifically, at $t\approx$11 Myr, the pair reaches its absolute closest approach with a minimum 3D intercluster separation of $\Delta_{\rm 3D}\approx$ 26 pc. This projected proximity, which is comparable to the sum of their tidal radii, suggests a grazing encounter that may lead to transient dynamical interactions or even a brief bound phase, highlighting the importance of considering future orbital evolution in the study of potential BOCs.

Although our results show that NGC~869 and NGC~884 are currently not gravitationally bound, as their present-day separation (103~pc) significantly exceeds the sum of their tidal radii ($\sim$26~pc), the forward integration of their orbits reveals a particularly intriguing outcome. After approximately 11 Myr, the 3D separation between the two OCs decreases to 26 pc, and is shown in Figure~\ref{fig:births}c. This future convergence suggests a potential close encounter or even a dynamical interaction event between the two OCs. Unlike the ongoing collision scenario of two unrelated clusters reported by \citet{Piatti_2022}, where IC \,4665 and Collinder \,350 first come into contact at the present epoch after forming independently, our system illustrates the inverse case, a pair of clusters that may experience significant mutual interaction in the future, despite not being currently bound. Such predictions highlight the dynamical richness of binary or near-BOCs and raise the possibility that gravitationally unbound cluster pairs can still engage in tidal encounters over long time-scales, depending on their relative orbital configurations within the Galactic potential.

\section{Dynamical Parameters and Mass Distribution}\label{Evolving_Times}
\subsection{Luminosity Function}

\begin{figure}[h]
\centering
\includegraphics[width=0.5\linewidth]{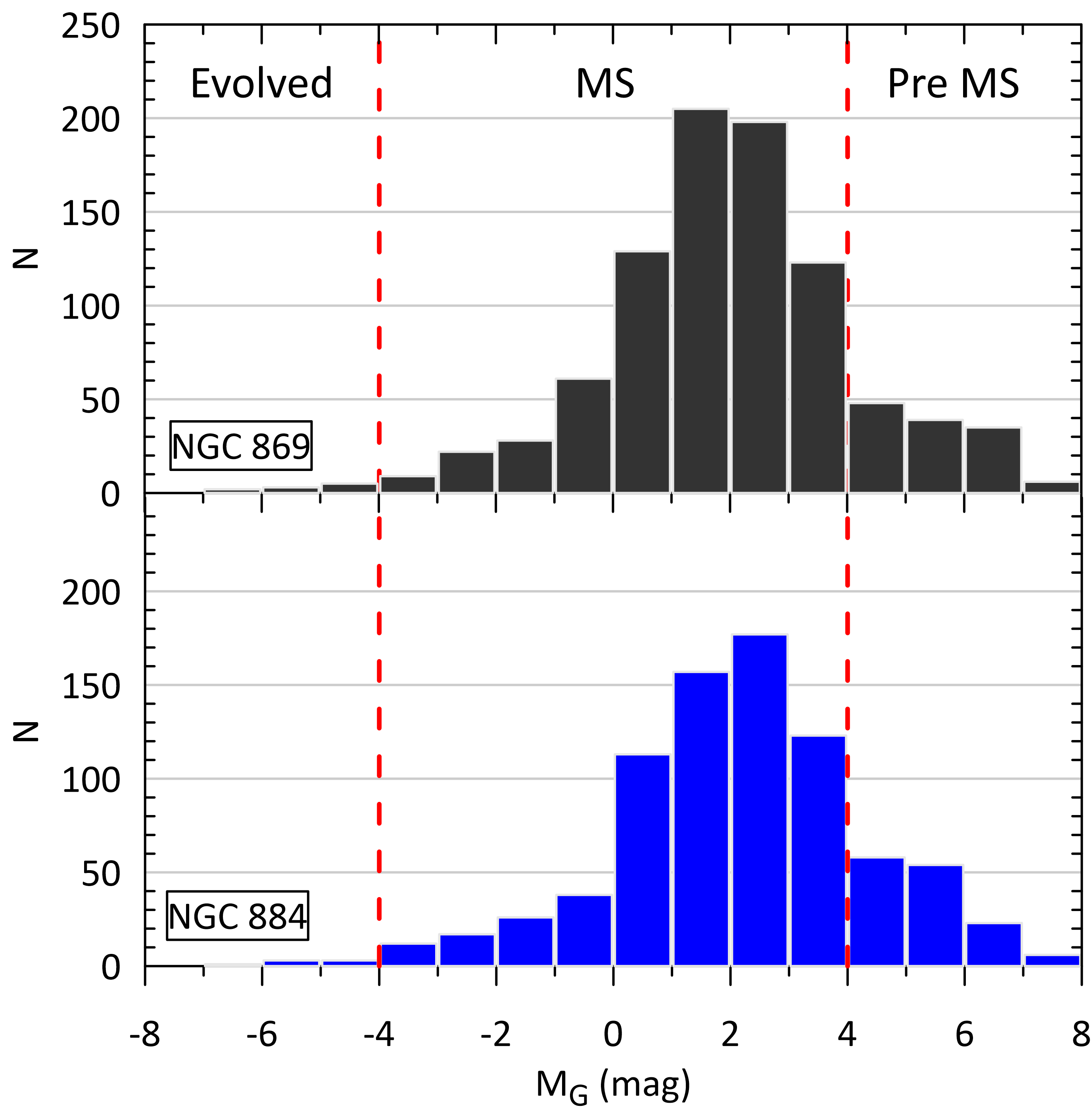}
\caption{LFs of cluster member stars belonging to different luminosity classes in NGC\,869 (upper panel) and NGC\,884 (lower panel).}
\label{LF}
\end{figure}

The stellar content of the BOC system NGC\,869 and NGC\,884 was further analyzed through the construction of luminosity and present-day mass functions (PDMFs) based on high-precision astrometric and photometric data from the \textit{Gaia}~DR3 release. The luminosity function (LF) characterises the distribution of stars over intervals of absolute $G$-band absolute magnitudes ($M_{\rm G}$). For this purpose, stars with reliable photometric and astrometric measurements were selected, and the LFs were computed in bins of 1 mag, which primarily trace the MS population. Figure~\ref{LF} displays the resulting LFs, indicating a relative incompleteness toward the faint end due to photometric limitations, particularly fainter than $G \sim 20.5$~mag. This observational bias impacts the detection of low-mass members and slightly alters the slope at the faint tail of the LF. The observed variations in the luminosity function, when considered together with previously reported radial trends in the $h$–$\chi$ Persei system, are consistent with mass segregation, wherein more massive stars are preferentially concentrated toward the cluster centers \citep[e.g.,][]{Slesnick02, Sharma08, Hasan11}.

\subsection{Present Day Mass Function}
\begin{figure}
\centering
\includegraphics[width=0.5\linewidth]{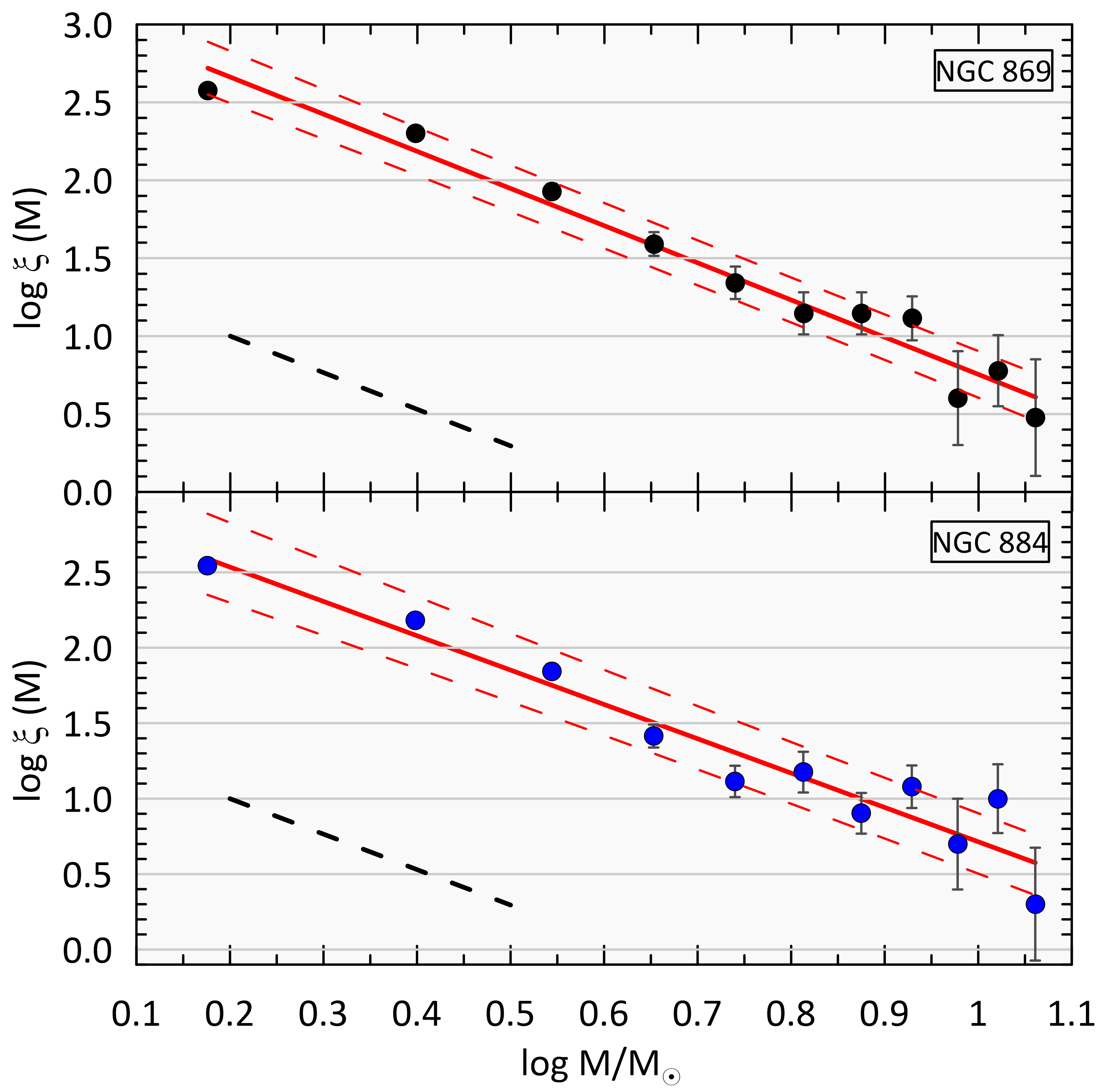}
\caption{PDMF of NGC\,869 (upper panel) and NGC 884 (lower panel). Red lines represent the PDMFs of both OCs, while red dashed lines are the $\pm 1\sigma$ standard deviations. The black dashed lines in the panels represent the slope of \citet{sal55}.}
\label{MF}
\end{figure}

To calculate the PDMFs of each OC, we adopted a standard power-law form of the initial mass function (IMF) expressed as follows:
\begin{equation}
\log \left(\frac{dN}{dM}\right) = -(1+\Gamma) \times \log(M) +\text{constant}
\label{eq:mf}
\end{equation}
{\normalsize
where $dN/dM$ represents the number of stars per unit mass interval and $\Gamma$ is the slope of the function. For stars with masses greater than $1\,M_{\odot}$, this formalism is consistent with the classical Salpeter IMF, which assumes $\Gamma = 1.35$ \citep{sal55}. Masses of the main-sequence stars were estimated using a fourth-order polynomial relation between $M_{\rm G}$ and $M/M_{\odot}$, calibrated using {\sc PARSEC} isochrones by \citet{Bressan_2012}, as follows:}

\begin{eqnarray}
\Biggr[\frac{M}{M_{\odot}}\Biggr]= 3.95592 - 1.57215\,M_{\rm G} + 0.25287\,M_{\rm G}^2 \\ \nonumber
+ 0.01953\,M_{\rm G}^3 - 0.00753\,M_{\rm G}^4.
\end{eqnarray}
Using this mass-luminosity relation (MLR), the masses of individual member stars were computed, and the resulting mass distributions were used to build the PDMFs of both OCs. Figure~\ref{MF} shows the derived PDMFs, along with their best-fit slopes. For NGC\,869, the slope is $\Gamma = 1.12 \pm 0.06$, and for NGC\,884, it is $\Gamma = 1.17 \pm 0.06$. Both values are in close agreement with the canonical \citet{sal55} slope as $\Gamma = 1.35$, suggesting that the stellar populations of the clusters follow a standard IMF for the mass range analysed. The relatively steep slopes also indicate an excess of low-mass stars compared to high-mass stars, as expected for young OCs.

\begin{table}
\centering
\caption{Stellar mass distribution of NGC\,869 and NGC\,884 subdivided by luminosity class (LC): evolved stars, main sequence (MS), and pre-main sequence (Pre-MS) members. $N$ is the number of stars in the luminosity class; $\sum M$ and $\langle M \rangle$ represent the total and mean masses, respectively, in solar masses.}
\renewcommand{\arraystretch}{0.8}
\setlength{\tabcolsep}{10pt}
\begin{tabular}{l|ccc}
\toprule
LC & \multicolumn{3}{c}{NGC\,869}\\
\cline{2-4}
 & $N$ & $\sum M (M_\odot)$ & $\langle M \rangle (M_\odot$) \\
\hline
Evolved & 9  & 103  & 11.44\\
MS      & 718 & 1940 & 2.70 \\
Pre-MS  & 81 & 76  & 0.94 \\
\hline
Total   & 808 & 2119 & 2.62 \\
\bottomrule
LC & \multicolumn{3}{c}{NGC\,884}\\
\cline{2-4}
 & $N$ & $\sum M (M_\odot$) & $\langle M \rangle (M_\odot$) \\
\midrule
Evolved & 7   & 80   & 11.43 \\
MS      & 606 & 1630 & 2.69 \\
Pre-MS  & 94 & 90  & 0.96  \\
\midrule
Total   & 707 & 1800 & 2.55  \\
\bottomrule
\end{tabular}
\label{tab:total_mass}
\end{table}

In the analysis of the mass distributions of both OCs investigated in this study, the luminosity classes of the cluster members were considered. The luminosity classes of the member stars were determined using the CMDs of the OCs, and the number of stars in these groups, their total masses, and mean masses were calculated and listed in Table~\ref{tab:total_mass}. As a result of the analysis, the total masses of NGC 869 and NGC 884 were calculated as 2119 $M/M_{\odot}$ and 1800 $M/M_{\odot}$, respectively. These findings indicate that NGC\,869 is more massive than NGC\,884.

\subsection{The Dynamical State of Mass Segregation}
The dynamical relaxation time ($T_\mathrm{relax}$) provides a crucial measure of the time required for a stellar system to approach a state of kinetic equilibrium, where energy redistribution among member stars leads to the stabilisation of internal motions. In the context of OCs such as NGC\,869 and NGC\,884, this timescale serves as a key diagnostic for assessing the level of internal dynamical evolution. A relatively short $T_\mathrm{relax}$ implies that the system has undergone considerable internal interactions such as two-body encounters and mass segregation, leading toward dynamical maturity. Conversely, a long relaxation time suggests that the cluster remains dynamically young and may still retain features reflecting its initial conditions.

To estimate $T_\mathrm{relax}$ for both OCs, we adopt the formalism introduced by \citet{Spitzer1971}, given as:
\begin{equation} \label{Eq:6}
T_\mathrm{relax} = \frac{8.9 \times 10^5 N^{1/2} R_\mathrm{h}^{3/2}}{\langle M_\mathrm{C} \rangle^{1/2} \log(0.4N)},
\end{equation}
where $N$ represents the total number of cluster members, $\langle M_\mathrm{C} \rangle$ is the mean stellar mass in Solar units, and $R_\mathrm{h}$ is the half-mass radius in parsecs. To derive $R_\mathrm{h}$, we apply the semi-empirical relation provided by \citet{Sableviciute_2006}:
\begin{equation}
R_\mathrm{h} = 0.547 \times r_\mathrm{c} \times \left( \frac{r_\mathrm{t}}{r_\mathrm{c}} \right)^{0.486},
\end{equation}
in which $r_\mathrm{c}$ and $r_\mathrm{t}$ correspond to the core and tidal radii, respectively. As mentioned in Section~\ref{section_structural}, the $r_\mathrm{t}$ parameter in the relation has been adopted as $r_\mathrm{lim}$. In this study, instead of estimating the tidal radius using an analytical expression, the values listed in Table~\ref{tab:rdp} were adopted. The core radii of NGC\,869 and NGC\,884 are obtained from the fitted RDPs, and using these parameters, we calculate the half-mass radii to be approximately $2.76 \pm 0.05$ pc for NGC\,869 and $3.54 \pm 0.05$ pc for NGC\,884. Substituting these values into Equation~\ref{Eq:6}, we derive the relaxation times as $T_\mathrm{relax} \approx 29$ Myr for NGC\,869 and $40$ Myr for NGC\,884. Since both OCs have ages of about 20 Myr, they have not yet completed a full relaxation cycle. Nevertheless, the presence of early mass segregation, common in dynamically young OCs, may reflect rapid internal evolution driven by their initial density structure or by primordial mass segregation, rather than long-term two-body relaxation.

Based on the \texttt{MCMC}-based distance estimates, the current separation between NGC\,869 and NGC\,884 is approximately 103 pc. The tidal radii of both OCs are calculated as 13.24~pc and 12.90~pc, respectively, yielding a total tidal extent of 26.14 pc. The ratio of their spatial separation to the combined tidal radius is therefore:
\begin{equation}
    \frac{d_{\rm separation}}{r_{\rm tidal,~total}} = \frac{103}{26.14} \approx 3.94.
\end{equation}

Since this ratio significantly exceeds unity, it indicates that both OCs lie quite beyond the tidal influence of each other. Consequently, we conclude that NGC\,869 and NGC\,884 are currently not gravitationally bound. Despite their close spatial proximity and similar ages, the current separation implies that they are a coeval but unbound OC pair, likely originating from a common star-forming region but no longer dynamically coupled.

The long-term dynamical evolution of OCs is significantly influenced by internal gravitational interactions among member stars, which lead to the gradual redistribution of energy and the eventual loss of members. One of the key timescales characterizing this process is the evaporation time ($\tau_{\rm ev}$), which represents the time required for the cluster to lose the majority of its stars due to internal relaxation mechanisms \citep{2001ApJ...553..744A}. For clusters in a state of virial equilibrium, the evaporation time is approximately $\tau_{\rm ev} \sim 100\times T_{\rm relax}$, where $T_{\rm relax}$ is the dynamical relaxation time. For the clusters studied here, this relation yields $\tau_{\rm ev} \approx 2000$ Myr, substantially longer than the canonical 100 Myr value cited for classical OCs. This demonstrates that their full dynamical dissolution proceeds on multi-Gyr timescales. To evaluate the current dynamical state of the clusters, we computed the dynamical evolution parameter $\tau = \mathrm{age}/T_{\rm relax}$, a dimensionless indicator of dynamical maturity, where systems with $\tau \gg 1$ are dynamically evolved or relaxed, and those with $\tau \ll 1$ are dynamically young and still relaxing. For NGC\,869 and NGC\,884, the derived $\tau$ values show that both OCs have advanced substantially in their dynamical evolution, indicating partial relaxation despite their relatively young ages.

\begin{figure}
\centering
\includegraphics[width=0.6\linewidth]{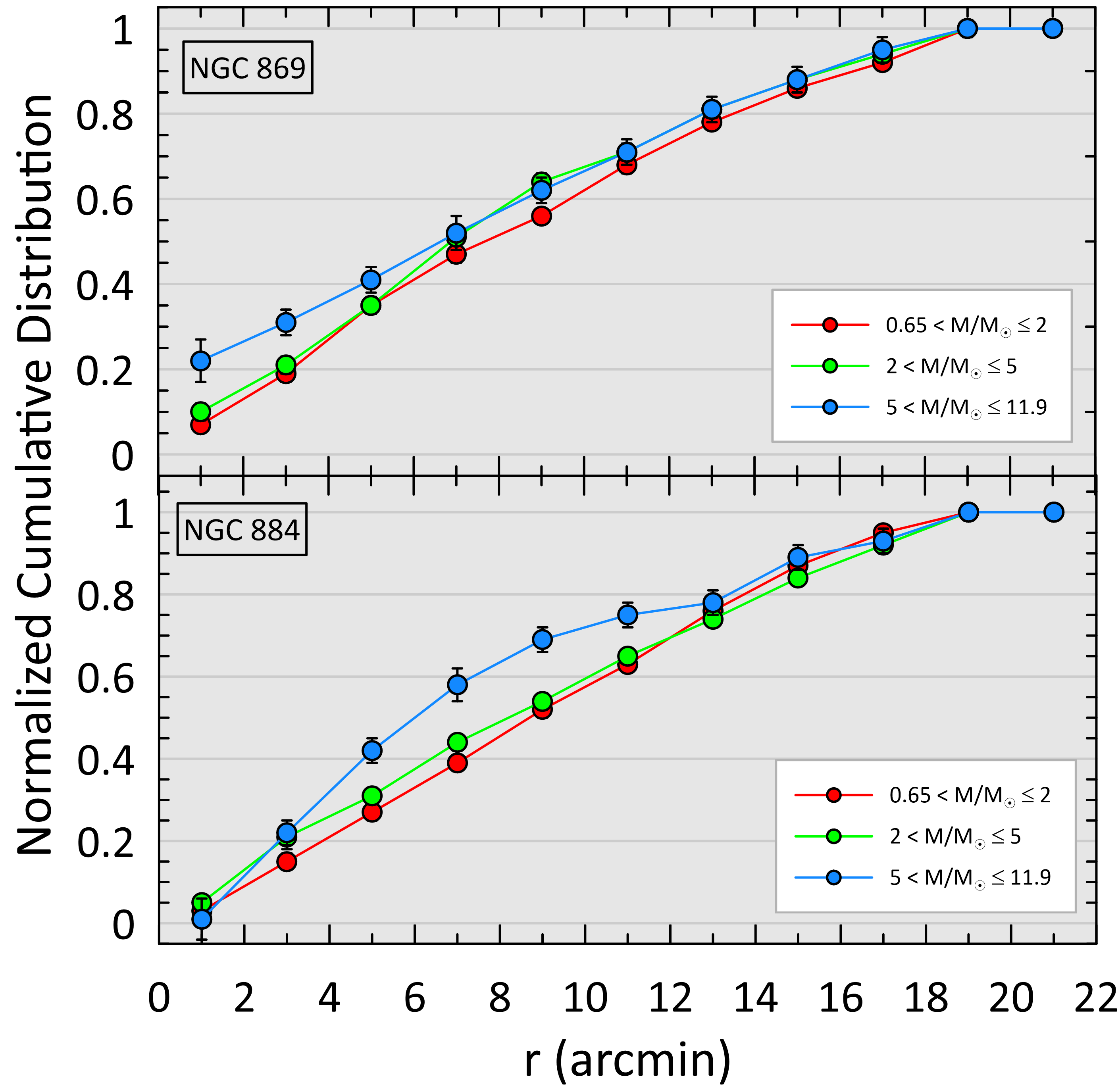}
\caption{Cumulative radial distributions of stars stratified by mass for NGC\,869 (upper panel) and NGC\,884 (lower panel).}
\label{fig:mass-seg}
\end{figure}

To examine mass segregation in the BOCs, cluster members were divided into three equal-number mass bins: high-mass ($5 < M/M_{\odot} \leq 11.9$), intermediate-mass ($2 < M/M_{\odot} \leq 5$), and low-mass ($0.65 < M/M_{\odot} \leq 2$) stars. The normalised cumulative radial distributions (Figure~\ref{fig:mass-seg}) reveal a clear mass segregation signature in both OCs, with high-mass stars preferentially concentrated toward the cluster cores, likely due to dynamical relaxation, while intermediate- and low-mass stars exhibit more extended radial distributions toward the outskirts. The results reveal a clear signature of mass segregation, manifested by the enhanced central concentration of high-mass stars relative to their lower-mass counterparts. Such a distribution is commonly interpreted as a consequence of dynamical relaxation, where two-body interactions drive more massive stars toward the cluster core over time. In contrast, intermediate- and low-mass stars exhibit broader radial distributions, preferentially populating the outer regions of the clusters. However, given the young ages of NGC\,869 and NGC\,884, the observed degree of mass segregation may not be fully explained by dynamical evolution alone. An alternative or complementary explanation is the presence of primordial mass segregation, whereby massive stars form preferentially in the densest central regions during the early stages of cluster formation \citep{Bonnell1998, Hillenbrand1998, Allison2009}. Several theoretical and observational studies have shown that massive stars can be centrally concentrated at birth, particularly in clusters formed within dense molecular environments, without requiring significant dynamical evolution \citep{McMillan2007, Parker2016}. Therefore, the mass segregation observed in the $h$ and $\chi$ Persei system is likely the result of a combination of early formation conditions and subsequent dynamical relaxation, consistent with their interpretation as co-evolved BOCs.

To quantitatively evaluate the degree of mass segregation, the Kolmogorov-Smirnov (K-S) statistical test was applied in a pairwise manner to compare the cumulative radial distributions of high-mass stars with those of intermediate- and low-mass stellar populations separately. For NGC\,869, the comparison between high-mass and intermediate-mass stars resulted in a $p$-value of 0.005, while the comparison between high-mass and low-mass stars yielded a $p$-value of 0.007. Similarly, for NGC\,884, the corresponding $p$-values were 0.003 and 0.009, respectively. In all cases, the obtained $p$-values are below 0.01, allowing us to reject the null hypothesis that the compared stellar populations are drawn from the same parent distribution. These statistically significant differences indicate that high-mass stars exhibit a spatial distribution distinct from that of lower-mass stars, providing strong evidence for the presence of mass segregation in both OCs under study.

\section{Discussion}\label{Discussion}

We have conducted a comprehensive investigation of the structural, astrometric, photometric, kinematical, and dynamical characteristics of the BOCs NGC\,869 and NGC\,884 using {\it Gaia} DR3 data. Our findings reinforce the well-established binary nature of these prominent BOCs, commonly known as $h$ and $\chi$ Persei \citep{Currie2010}. While their physical association has been widely acknowledged in the literature, a unified and multi-dimensional analysis encompassing all major cluster properties has been lacking. This study fills that gap by presenting an integrated approach that brings together multiple lines of evidence, offering a more complete understanding of their formation history and dynamical state.

Despite the lack of spectroscopic data for both OCs, the agreement between the mean values calculated from the distributions of the parameters obtained by the \texttt{MCMC} and SED methods revealed that the fundamental parameters for both OCs were determined with high accuracy and precision. The fundamental parameters ($T_{\rm eff}$, $\log g$, [Fe/H], $A_{\rm V}$ and $d$) were determined from the SED analyses of total 635 member stars in both OCs, and the histograms of the parameters were plotted and their mean values were calculated by fitting Gaussian curves. Both OCs exhibit slightly metal-poor metallicities of $[{\rm Fe/H}] = -0.25 \pm 0.03$ dex and consistent mean $V$-band extinctions of $A_{\rm V} \approx 1.5$ mag, corresponding to colour excess values of $E(B-V) \approx 0.48$ mag. The distance distributions are around $2300$ pc, consistent with previous results based on isochrone fitting and \textit{Gaia} trigonometric parallaxes \citep[e.g.,][]{Currie2010, Cantat-Gaudin2018}. 

A critical challenge in SED-based stellar characterisation lies in the intrinsic parameter degeneracies, most notably among effective temperature, $V$-band extinction, and stellar radius, that can bias the physical modeling \citep{McDonald2017}. To address this, we applied a multi-faceted approach aimed at reducing degeneracy and driven uncertainty. Our method incorporated high-precision \textit{Gaia} DR3 trigonometric parallaxes, extinction priors from 3D Galactic dust maps \citep{Schlafly2011}, and Bayesian model averaging across two independent stellar atmosphere model grids (PHOENIX and Castelli-Kurucz). Additionally, the use of nested sampling with high posterior resolution allowed for better exploration of the multi-dimensional parameter space. These efforts yielded robust parameter estimates for both OCs with minimal scatter. The reduction of model-dependent biases supports our conclusion that these clusters form a physically bound system \citep{bragg05}. Moreover, these consistent parameter values suggest that both OCs formed under very similar interstellar conditions within the Galactic disc \citep{Fujimoto1996, Fellhauer2009}.

Dynamical orbit integration using high-precision astrometric data from \textit{Gaia} DR3 reveals that both OCs exhibit remarkably similar Galactic trajectories and nearly indistinguishable birthplaces, providing strong kinematic support for their coeval formation. The orbital paths traced over the past about $t=20$ Myr show minimal divergence in space velocities and proper-motion components within the Galactic disc, effectively ruling out scenarios of gravitational capture or chance alignment. Instead, the results support a primordial BOC origin, in which both OCs were born simultaneously in closely related regions of the Galactic disc. Furthermore, both OCs are currently not gravitationally bound due to their large separation relative to their tidal radii, orbital integration predicts a close approach in $\approx$11 Myr. \textcolor{black}{At that time, the two systems are expected to experience a closest approach of approximately 26 pc, which may indicate a future dynamical interaction.} This contrasts with scenarios where clusters collide after independent formation, highlighting instead a case where initially unbound OCs may engage in significant tidal encounters over time. These results underline the complex dynamical evolution of binary or near-BOCs and the potential for gravitational interactions long after their formation \citep{Piatti_2022}.

Both OCs exhibit similar total stellar masses, estimated as $2119\,M_\odot$ for NGC\,869 and $1800\,M_\odot$ for NGC\,884. The slopes of their PDMFs also show a close match, with $\Gamma = 1.12 \pm 0.06$ and $1.17 \pm0.06$, respectively, closely aligning with the canonical Salpeter value. Using the isochrone fitting method, BOC ages are obtained as $\log (t/{\rm yr})\cong 7.30\pm 0.15$ for both OCs. The same age estimates, combined with the detection of mild mass segregation in both OCs cores, suggest a shared dynamical evolution \citep{Horie24}. Taken together, these findings support a scenario in which NGC\,869 and NGC\,884 formed nearly simultaneously in closely related regions of the Perseus arm. The formation could have been triggered by large-scale processes such as a supernova shock wave \citep{Boss12, Iffrig15} or a cloud-cloud collision \citep{Horie24}, both of which are consistent with models of BOC formation in dense stellar environments \citep{slesnick_2002}.

The CMDs of both OCs exhibit well-defined and nearly identical main-sequence and pre-main-sequence stellar populations \citep{slesnick_2002}. The \texttt{MCMC}-based isochrone fitting and SED analyses used to determine the fundamental parameters yield consistent metallicity and extinction estimates. The alignment of the evolutionary sequences in both OCs further suggests that NGC\,869 and NGC\,884 formed within closely related regions of the Perseus spiral arm \citep{Reid_2019, Currie2010}. Taken together, these multi-dimensional diagnostics indicate that both OCs are not only spatially and kinematically associated, but also share a common evolutionary history. The combined photometric, kinematic, and orbital evidence supports the interpretation of NGC\,869 and NGC\,884 as primordial BOCs within the Perseus arm. Nevertheless, uncertainties remain regarding their interaction with the surrounding OB association (Per~OB1), and future multi-wavelength and spectroscopic studies will be essential to better constrain the environmental effects that may have influenced their subsequent evolution \citep{Cappa_1996}.

Ultimately, the binary nature of NGC\,869 and NGC\,884 offers a unique laboratory for studying the early dynamical evolution of massive clusters, the impact of the Galactic tidal field, and the mechanisms by which bound cluster pairs survive over tens of millions of years. Continued high-precision astrometry, combined with advanced stellar population modeling and N-body simulations, will be essential to fully clarify the formation pathways and dynamical lifetimes of such BOCs. Future high-precision astrometric and spectroscopic studies could provide tighter constraints on age spreads, binarity, and internal dynamics, offering deeper insight into the co-evolution of these closely interacting clusters.

\normalem
\begin{acknowledgements}
We thank the anonymous referee for the constructive comments and suggestions that helped improve the quality of this manuscript. This study has been supported in part by the Scientific and Technological Research Council (T\"UB\.ITAK) 122F109. This work presents results from the European Space Agency (ESA) space mission $Gaia$. $Gaia$ data are being processed by the $Gaia$ Data Processing and Analyzis Consortium (DPAC). Funding for the DPAC is provided by national institutions, in particular, the institutions participating in the $Gaia$ Multi-Lateral Agreement (MLA). The $Gaia$ mission website is \url{https://www.cosmos.esa.int/gaia}. The $Gaia$ archive website is \url{https://archives.esac.esa.int/gaia}. The authors would like to express their gratitude to the Deanship of Scientific Research at Northern Border University, Arar, KSA, for funding this research under project number "NBU-FFR-2026-237-02".
\end{acknowledgements}
  
\bibliographystyle{raa}
\bibliography{ms2025-0422}

\end{document}